\documentclass{article}
\usepackage[english]{babel}
\usepackage[round]{natbib} 
\usepackage{amssymb}
\usepackage{amsmath}
\usepackage{bbm}
\usepackage{algorithm}
\usepackage{hyperref}
\usepackage{graphicx}
\usepackage{dsfont}
\usepackage{cancel}
\usepackage{url}
\usepackage{algpseudocode}
\usepackage[a4paper,
            bindingoffset=0.2in,
            left=1in,
            right=1in,
            top=1in,
            bottom=1in,
            footskip=.25in]{geometry}

\title{\textbf{Bayesian inference for spatio-temporal hidden Markov models using the exchange algorithm}}
\author{Daniele Tancini$^1$, Riccardo Rastelli$^2$ and Francesco Bartolucci$^1$}
\date{$^1$Department of Economics, University of Perugia, Italy\\
$^2$School of Mathematics and Statistics, University College Dublin, Ireland}

\begin{document}

\maketitle

\section*{Abstract}
Spatio-temporal hidden Markov models are extremely difficult to estimate because their latent joint distributions are available only in trivial cases. In the estimation phase, these latent distributions are usually substituted with pseudo-distributions, which could affect the estimation results, in particular in the presence of strong dependence between the latent variables. In this work, we propose a spatio-temporal hidden Markov model where the latent process is an extension of the autologistic model. We show how inference can be carried out in a Bayesian framework using an approximate exchange algorithm, which circumvents the impractical calculations of the normalizing constants that arise in the model. Our proposed method leads to a Markov chain Monte Carlo sampler that targets the correct posterior distribution of the model and not a pseudo-posterior. In addition, we develop a new initialization approach for the approximate exchange method, reducing the computational time of the algorithm. An extensive simulation study shows that the approximate exchange algorithm generally outperforms the pseudo-distribution approach, yielding more accurate parameter estimates. Finally, the proposed methodology is applied to a real-world case study analyzing rainfall levels across Italian regions over time.

\vspace{0.8cm}

\noindent \textbf{Keywords:} Exchange algorithm, Hidden Markov model, Intractable normalizing constant, Spatio-temporal model

\section{Introduction}\label{sec:int}
In the era of complex data, where information is linked to both spatial and temporal dimensions, spatio-temporal models offer a powerful framework for capturing and analyzing dynamic patterns. In this context, spatio-temporal hidden Markov (HM) models have been used in diverse applications, including the analysis of COVID-19 \citep{bartolucci2022spatio}, global food accessibility \citep{bartolucci2022hidden}, and crime data \citep{robertson2022predicting}. However, a key challenge of these methodologies lies in the intractability of the distribution for the latent variables, which makes inference on the model parameters computationally demanding or proibitive.
For this reason, in the estimation phase the distribution of the unobservable component is usually replaced by a pseudo-distribution; see \cite{besag1974spatial}.
However, there are concerns that this approximation could affect the precision of estimates, presumably when the model presents a high degree of complexity.
In this paper, we address this issue by proposing a new computational framework to perform inference on a class of Bayesian spatio-temporal HM models. Our algorithm does not rely on pseudo-distributions and it delivers a sample from the posterior distribution of the model using a variant of the exchange algorithm.

In a Bayesian setting, the idea of a pseudo-posterior distribution \citep{besag1974spatial} has been used in various works, including \cite{bouranis2017efficient}, where also a calibration method for the pseudo-distribution was proposed. In the context of spatio-temporal HM models, the same approach is usually followed, replacing an intractable distribution of the latent variables with a pseudo-distribution \citep{bartolucci2022hidden,bartolucci2022spatio}. Alternative solutions to address the computational intractability previously described are based on methods which do not require to evaluate the likelihood function at all, such as approximate Bayesian computation \citep{marin2012approximate} and Bayesian synthetic likelihoods \citep{price2018bayesian}. A more specific type of intractability 
may be limited to 
the normalizing constants associated to a particular distribution, or likelihood function. In such cases, possible solutions are given by the single auxiliary variable method \citep{moller2006efficient} and the exchange algorithm \citep{murray2012mcmc}.

The exchange algorithm has been extensively used in the context of exponential random graph models (ERGM) \citep{caimo2011bayesian} and more in general for so-called doubly intractable problems \citep{murray2012mcmc}. A noisy variant, meaning that it approximates the target distribution, 
of the algorithm can be found in \cite{alquier2016noisy}, and other relevant extensions have been proposed by \cite{liang2010double}, \cite{lyne2015russian} and \cite{liang2016adaptive}. More recently, \cite{yuan2024markov} have proposed a novel idea in the context of doubly-intractable distributions, whereby the authors introduce auxiliary variables both in the proposals and in the acceptance–rejection step. The reader can refer to \cite{park2018bayesian} for an extensive review on Bayesian inference in the presence of intractable normalizing constants.

In this work, we consider a general spatio-temporal hidden Markov model, where the latent process follows an autologistic model \citep{besag1974spatial} characterized by an intractable normalizing constant. Our latent variable framework generalizes the autologistic model to a $K$-state process, building upon and extending the models introduced by the recent works of \cite{bartolucci2022hidden,bartolucci2022spatio}.
Our new structure includes sets of parameters that characterize the prevalence of each of the latent states, as well as their spatial and temporal dependencies. The model parameters characterize separately the initial state of the system using a dedicated set of parameters. Conditionally on the HM latent structure, the distribution of the observed data can be defined in full generality thus ensuring wide applicability of the methodology.

A central contribution of our work relates to model inference and computation. We move away from previous approaches based on pseudo-posteriors and instead adopt a variant of the exchange algorithm, which gets embedded within a Gibbs sampler framework. The resulting sampler targets the correct posterior distribution and we show that it leads to more accurate inference than other available methods.
Specifically, our algorithm is an approximate exchange algorithm \citep{friel2011classification,caimo2011bayesian}, whereby data augmentation is used on the latent variables, thus creating auxiliary variables. In the original exchange algorithm, an exact simulation of the auxiliary variables is required in order to simplify the acceptance rate of the Metropolis-Hastings (MH) scheme \citep{metropolis1953equation,hastings1970monte}.
The approximate exchange algorithm does not sample the auxiliary from a perfect simulator, but rather it uses a Gibbs sampler to obtain a random draw, which is taken from the last iteration of such auxiliary sampler. A theoretical justification for the validity of this approach is provided in \cite{everitt2012bayesian}.

In our framework, the auxiliary variables consists of an auxiliary spatio-temporal process that follows the same distribution of the latent process.
However, using the approximate exchange algorithm for spatio-temporal HM models can be computationally intensive, because the number of iterations for the auxiliary variables increases with the model complexity. This problem has been studied in \cite{bhamidi2008mixing}, where the authors show that convergence, when sampling from a very large ERGM through MCMC, is likely to be slow. The same authors also suggest that one should take a conservative approach and choose a large number of auxiliary iterations. However, as a consequence, the resulting exchange algorithm may be computationally infeasible for large graphs. To manage this computational problem, we propose a new initialization strategy for the auxiliary process within the approximate exchange algorithm: we impose that the distribution over the auxiliary variables must be equal to the distribution of the latent component for each initialization of the auxiliary Gibbs sampler. Specifically, we use the latent variables from the previous iteration as the starting values for the auxiliary process. This aims at dramatically reducing the number of iterations required before reaching a suitable draw for the auxiliary variables. This choice is motivated by the expectation that, due to the sequential update of each parameter, the latent state from the previous iteration lies closer to the target distribution of the auxiliary process.

To validate our proposed methodology, we compare the pseudo-posterior approach and our algorithm in a broad simulation study, showing that our approximate exchange solution provides more accurate estimates within a reasonable computational time. Finally, we conclude showing an application of the proposed model to the analysis of meteorological trends in Italy, focusing specifically on regional-level precipitation data.

The reminder of the paper is organized as follow. In Section \ref{sec:mod} we describe the class of models proposed, focusing in particular to the models with Gaussian responses. In Section \ref{sec:bayes} we discuss the Bayesian estimation of the model. 
In Section \ref{sec:sim} we discuss a simulation study, which compares the pseudo-posterior approach and the approximate exchange algorithm. Finally, we conclude with Section \ref{sec:apl} considering a real data application.

\section{Spatio-temporal hidden Markov models}\label{sec:mod}

In this section we propose a general spatio-temporal hidden Markov model that extends the models proposed in \cite{bartolucci2022hidden} and \cite{bartolucci2022spatio}, and we discuss the interpretability of the model. The main differences between the previous two models and the proposed one are emphasized in the following section. Finally, focusing on a version where the response variable is a (multivariate) Gaussian distribution, we describe the complete model.

\subsection{Model}\label{sec:model}

Let $\mathcal{S} \subset \mathbb{N}$ be the site space set and $\mathcal{T} \subset \mathbb{N}$ be the time occasion set, given a suitable probability space, we consider 
\begin{equation*}
\{\boldsymbol{Y}_{i,t}, U_{i,t} : (i,t) \in \mathcal{S} \times \mathcal{T} \},
\label{eq:proc}
\end{equation*}
with $\boldsymbol{Y}_{i,t} \in \mathcal{Y} \subseteq \mathbb{R}^d$, $d \in \mathbb{N}$, and $U_{i,t} \in \mathcal{U} \subseteq \mathbb{N}$. 
In our formulation $\mathcal{U} = \{1,\ldots,K\}$, where $K$ denotes the number of states associated to the latent process. 
In practice, this number is usually chosen according to an information criterion. However, it may also be possible to treat $K$ as a random variable.

A general spatio-temporal hidden Markov model is defined as follows:
\begin{equation*}
\boldsymbol{Y}_{i,t} \lvert U_{i,t} = u \sim \mathcal{L}_u,
\label{eq:st}
\end{equation*}
where $\mathcal{L}_u$ is a probability distribution, which depends on $u$. We assume for 
$$\{U_{i,t}\} = \{U_{i,t} : (i,t) \in \mathcal{S} \times \mathcal{T} \}$$ a first order Markov (time) dependence combined with a Markov random property (space) with respect to (w.r.t.) a neighbourhood system $\mathcal{H}$. The neighbourhood system is equal for all $t \in \mathcal{T}$, and it is defined as $\mathcal{H} = \{\eta_i : i \in \mathcal{S} \}$, where $\eta_i$ is the neighborhood of site $i$ so that if $i \notin \eta_i$ and $j \in \eta_i$ then $i \in \eta_j$.
This means that 

\begin{equation}
p(U_{i,t} = k \lvert \boldsymbol{U}_{-(i,t)} = \boldsymbol{u}_{-(i,t)}, \boldsymbol{\theta}) = p(U_{i,t} = k \lvert 
\tilde{\boldsymbol{U}}_{i,t} = \tilde{\boldsymbol{u}}_{i,t}, U_{i,t-1} = u_{i,t-1}, \boldsymbol{\theta}),
\label{eq:dip}
\end{equation}
where $\boldsymbol{U}_{-(i,t)} = \boldsymbol{u}_{-(i,t)}$ stands for the vector of all $\boldsymbol{U}$ except for $U_{i,t}$, while $\tilde{\boldsymbol{U}}_{i,t} = \tilde{\boldsymbol{u}}_{i,t}$ defines the collection of the variables neighborhood of the variable at site $i$, that is, $\boldsymbol{U}_{j \in \eta_i,t} = \boldsymbol{u}_{j \in \eta_i,t}$. 

Let $\boldsymbol{\theta} \in \boldsymbol{\Theta} \subseteq \mathbb{R}^p$ be the collection of parameters of the distribution of $\{U_{i,t}\}$, having probability mass function expressed as

\begin{equation}
p(\boldsymbol{u} \lvert \boldsymbol{\theta}) = \frac{q_{\boldsymbol{\theta}} (\boldsymbol{u})}{\mathcal{Z}_{\boldsymbol{\theta}}},
\label{eq:ratio}
\end{equation}
where $q_{\boldsymbol{\theta}} (\boldsymbol{u}) = \exp{\left[\boldsymbol{f}(\boldsymbol{u})' \boldsymbol{\theta}\right]}$ and $\mathcal{Z}_{\boldsymbol{\theta}} = \sum_{\boldsymbol{u}} q_{\boldsymbol{\theta}} (\boldsymbol{u})$, the latter constant usually being impractical to calculate. This intractability arises since the sum is extended to all possible latent configurations $\boldsymbol{u}$, requiring a huge or proibitive computational effort. 

Starting from the autologistc model proposed by \cite{besag1974spatial}, we propose an extension to the spatio-temporal setting. In particular, following \cite{bartolucci2022hidden},
we consider $\mathcal{T} = \{1,\ldots,T\}$ and $\mathcal{S} = \{1,\ldots,N\}$, and assume the following form:

\begin{equation}
\begin{split}
\log{q_{\boldsymbol{\theta}}(\boldsymbol{u})} & = \sum_{i=1}^N \sum_{u=1}^{K-1} \mathds{1}(U_{i,1} = u)\beta_u + \sum_{i=1}^{N-1} 
\sum_{\substack{j = i+1 \\  j \in \eta_i}}^{N} \sum_{u=1}^{K} \sum_{\substack{v=1 \\ v \neq u}}^{K} \mathds{1}(U_{i,1} = u, U_{j,1} = v)\gamma_{u,v} \\
& + \sum_{t>1} \left[ \sum_{i=1}^N \sum_{u=1}^{K-1} \mathds{1}(U_{i,t} = u)\beta^*_u + \sum_{i=1}^{N-1} 
\sum_{\substack{j = i+1 \\  j \in \eta_i}}^{N} \sum_{u=1}^{K} \sum_{\substack{v=1 \\ v \neq u}}^{K} \mathds{1}(U_{i,t} = u, U_{j,t} = v)\gamma^*_{u,v}   \right. \\
& \left. + \sum_{i=1}^{N} \sum_{u=1}^{K} \sum_{\substack{v=1 \\ v \neq u}}^{K} \mathds{1}(U_{i,t-1} = u, U_{i,t} = v) \delta_{u,v}\right],
\end{split}
\label{eq:lat}
\end{equation}
where $\mathds{1}(\cdot)$ is an indicator function, $\boldsymbol{u} = \{u_{1,1}, \ldots, u_{N,T}\}$ is the collection of the realized latent variables, and $\beta_K = \beta_K^* = 0$, as well as $\gamma_{u,u} = \gamma^*_{u,u} = \delta_{u,u} = 0$ for $u=1,\ldots,K$. Notice that the model defined in Equation \eqref{eq:lat} satisfies Equation \eqref{eq:dip}; see Appendix A for details. 

In comparison to the model described in \cite{bartolucci2022hidden}, the new model includes specific spatial initial time ($t=1$) parameters, that are $\gamma_{u,v}$ for $u \neq v$. In addition, when $t>1$, specific parameters for the prevalence of single and transition-states parameters, which are $\beta_u^*$ for $u = 1,\ldots,K-1$, $\gamma_{u,v}^*$, and $\delta_{u,v}$, $u \neq v$, $u = 1,\ldots,K$, are introduced for both spatial and temporal components.

Similarly to \cite{bartolucci2022hidden}, a typical assumption of these models is that of local independence, meaning that observable 
vectors $\boldsymbol{Y}_{i,t}$ are conditionally 
independent given the latent variables $\{U_{i,t}\}$. This implies that

$$p(\boldsymbol{y} \lvert \boldsymbol{u}, \boldsymbol{\xi}) = \prod_{i=1}^N \prod_{t=1}^T p(\boldsymbol{y}_{i,t} \lvert u_{i,t}, \xi_{u_{i,t}}),$$
where $\boldsymbol{y} = \{\boldsymbol{y}_{1,1}, \ldots, \boldsymbol{y}_{N,T}\}$, $\boldsymbol{\xi} = (\xi_1, \ldots, \xi_K)' \in \boldsymbol{\Xi}\subseteq \mathbb{R}^K$ is the vector parameter of $\boldsymbol{Y}_{i,t}$, considering $K$ states for the latent variables. 

\subsection{Parameters interpretation}\label{sec:modint} 

The number of parameters in Equation \eqref{eq:lat} is on order of $2(K-1) + 3K(K-1)$. The parameters for the initial time ($t=1$) are collected in $\boldsymbol{\beta} = (\beta_1, \ldots, \beta_{K-1}, 0)'$ and 
$$\boldsymbol{\gamma} = \begin{pmatrix}
0 & \gamma_{1,2} & \cdots & \gamma_{1,K}\\
\gamma_{2,1} & 0  & \ddots & \vdots \\
\vdots & \ddots & 0 & \gamma_{K-1,K} \\
\gamma_{K,1} & \cdots & \gamma_{K,K-1} & 0\\
\end{pmatrix},$$
whose diagonal terms are set to zero. In particular, with $\boldsymbol{\beta}$ we denote the vector parameter of the prevalence of the single states, while $\boldsymbol{\gamma}$ is the matrix of spatial dependence. In practice, $\boldsymbol{\beta}$ describes the prevalence of a single state: if $\beta_1 > \beta_2$ then there is a higher probability of state 1 w.r.t. state 2. 
Regarding $\boldsymbol{\gamma}$, each $\gamma_{u,v}$ defines the spatial dependence among the state $u$ in the site $i$ and the state $v$ in the neighborhood of site $i$. 

For $t>1$, the model is parametrized by $\boldsymbol{\beta^*} = (\beta^*_1, \ldots, \beta^*_{K-1},0)'$, 

$$\boldsymbol{\gamma^*} = \begin{pmatrix}
0 & \gamma^*_{1,2} & \cdots & \gamma^*_{1,K}\\
\gamma^*_{2,1} & 0  & \ddots & \vdots \\
\vdots & \ddots & 0 & \gamma^*_{K-1,K} \\
\gamma^*_{K,1} & \cdots & \gamma^*_{K,K-1} & 0\\
\end{pmatrix},$$
and

$$\boldsymbol{\delta} = \begin{pmatrix}
0 & \delta_{1,2} & \cdots & \delta_{1,K}\\
\delta_{2,1} & 0  & \ddots & \vdots \\
\vdots & \ddots & 0 & \delta_{K-1,K} \\
\delta_{K,1} & \cdots & \delta_{K,K-1} & 0\\
\end{pmatrix},$$ 
where $\boldsymbol{\delta}$ is the matrix characterizing temporal dependencies. The interpretation of $\boldsymbol{\beta}^*$ and $\boldsymbol{\gamma}^*$ is the same as the previous one, with the main difference being that these parameters are specific for $t>1$. The matrix $\boldsymbol{\delta}$ denotes the temporal dependencies between the random variable for site $i$ and time $t$ and the same site at time $t-1$. Note that none of these parameters vary over time; however, potential extensions could be considered in this regard.

We denote by $\boldsymbol{\theta} = \{ \boldsymbol{\beta}, \boldsymbol{\beta^*}, \boldsymbol{\gamma}, \boldsymbol{\gamma^*}, \boldsymbol{\delta} \}$ the collection of parameters of the distribution of $\{U_{i,t}\}$. Note that matrices $\boldsymbol{\gamma}, \boldsymbol{\gamma^*}$, and $\boldsymbol{\delta}$ are in general not symmetric and imposing zeros on diagonals and on the last element of vectors $\boldsymbol{\beta}$ and $\boldsymbol{\beta^*}$, allow us to reduce the number of parameters, which helps with the identifiability of the model.
In addition, we note that, in Equation \eqref{eq:lat}, we impose sums over $j=i+1 : j \in \eta_i$ to avoid problems of identifiability related to pairwise edges. The example below illustrates a case where the extra constraints can avoid non-identifiability issues.

\vspace{0.2cm}
\noindent
\textit{Example 1.1.}

\vspace{0.2cm}

\begin{figure}[h]
\centering
\includegraphics[width=0.3\textwidth]{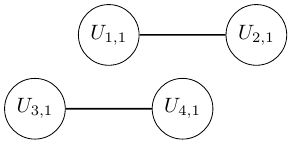} 
\caption{\textit{Graphical spatial dependence for the model defined in Example 1.1.} }
\label{figus}
\end{figure}

\noindent
\textit{Assume $N = 4$, focusing on $t = 1$, with only one edge between node 1 and 2, and other between node 3 and 4, as represented in Figure \ref{figus}.}

\noindent
\textit{Assume $U_{1,1} = k_1, U_{2,1}=k_2$ and $U_{3,1} = k_1, U_{4,1}=k_2$, where $k_1,k_2 \in \{1,\ldots,K\}$. If we consider the following form
$$\sum_{i=1}^{N-1} \sum_{j \in \eta_i} \sum_{u=1}^{K} \sum_{\substack{v=1 \\ v \neq u}}^{K} \mathds{1}(U_{i,1} = u, U_{j,1} = v)\gamma_{u,v},$$
we get
$$2\gamma_{k_1,k_2} + 2\gamma_{k_2,k_1} = 2(\gamma_{k_1,k_2} + \gamma_{k_2,k_1}),$$
implying that the map $\boldsymbol{\theta} \rightarrow p(\boldsymbol{u} \lvert \boldsymbol{\theta})$ is not one-to-one. In fact, the distribution of $\boldsymbol{u}$ depends on $\gamma_{k_1,k_2}$ and $\gamma_{k_2,k_1}$ only through their sum and so the two parameters are not identifiable. Using the form provided in Equation \eqref{eq:lat}, that is, $j=i+1 : j \in \eta_i$, we solve this problem. In addition, note that this issue may be resolved if we consider a symmetric matrix for $\boldsymbol{\gamma}$.} 

\vspace{0.5cm}

Since the number of parameters increases quadratically with $K$, we briefly describe possible parsimonious parameterizations. As starting point, we can consider only the upper triangular part of the matrices $\boldsymbol{\gamma}, \boldsymbol{\gamma^*}$, and $\boldsymbol{\delta}$, imposing a symmetric constraint. Following this parametrization, we are assuming same dependencies between spatial and temporal components from $u$ to $v$ and $v$ to $u$, where $u \neq v$ and $u,v = 1,\ldots,K$.  Moreover, we can assume that there is no difference between $t=1$ and $t >1$, that is, $\boldsymbol{\beta} = \boldsymbol{\beta^*}$ and $\boldsymbol{\gamma} =\boldsymbol{\gamma^*}$. 

A common way to interpret changes in the latent variables relies on the odds, defined as a ratio of conditional probabilities, of changing one latent variable while keeping all others fixed. We now proceed to characterize these odds for our model. Let us define a baseline level $k \in \mathcal{U}$. First, note that 

\begin{equation}
\begin{split}
\frac{p(U_{i,t} = w \lvert \boldsymbol{U}_{-(i,t)} = \boldsymbol{u}_{-(i,t)}, \boldsymbol{\theta})}{p(U_{i,t} = k \lvert \boldsymbol{U}_{-(i,t)} = \boldsymbol{u}_{-(i,t)}, \boldsymbol{\theta})} & = \frac{p(U_{i,t} = w, \boldsymbol{U}_{-(i,t)} = \boldsymbol{u}_{-(i,t)} \lvert \boldsymbol{\theta}) / p(\boldsymbol{U}_{-(i,t)} = \boldsymbol{u}_{-(i,t)} \lvert \boldsymbol{\theta})}{p(U_{i,t} = k, \boldsymbol{U}_{-(i,t)} = \boldsymbol{u}_{-(i,t)} \lvert \boldsymbol{\theta}) / p(\boldsymbol{U}_{-(i,t)} = \boldsymbol{u}_{-(i,t)} \lvert \boldsymbol{\theta})} \\
& = \frac{p(U_{i,t} = w, \boldsymbol{U}_{-(i,t)} = \boldsymbol{u}_{-(i,t)} \lvert \boldsymbol{\theta})}{p(U_{i,t} = k, \boldsymbol{U}_{-(i,t)} = \boldsymbol{u}_{-(i,t)} \lvert \boldsymbol{\theta})}\\
& = \frac{q_{\boldsymbol{\theta}}(U_{i,t} = w, \boldsymbol{U}_{-(i,t)} = \boldsymbol{u}_{-(i,t)})/ \mathcal{Z}_{\boldsymbol{\theta}}}{q_{\boldsymbol{\theta}}(U_{i,t} = k, \boldsymbol{U}_{-(i,t)} = \boldsymbol{u}_{-(i,t)})/ \mathcal{Z}_{\boldsymbol{\theta}}}\\
& = \frac{q_{\boldsymbol{\theta}}(U_{i,t} = w, \boldsymbol{U}_{-(i,t)} = \boldsymbol{u}_{-(i,t)})}{q_{\boldsymbol{\theta}}(U_{i,t} = k, \boldsymbol{U}_{-(i,t)} = \boldsymbol{u}_{-(i,t)})},
\end{split}
\label{eq:rat}
\end{equation}
for $w \in \mathcal{U}$, $w \neq k$.  
Taking the log$(\cdot)$ of both terms of Equation \eqref{eq:rat} we have

$$\log{\frac{p(U_{i,t} = w \lvert \boldsymbol{U}_{-(i,t)} = \boldsymbol{u}_{-(i,t)}, \boldsymbol{\theta})}{p(U_{i,t} = k \lvert \boldsymbol{U}_{-(i,t)} = \boldsymbol{u}_{-(i,t)}, \boldsymbol{\theta})}} =\log{\frac{q_{\boldsymbol{\theta}}(U_{i,t} = w, \boldsymbol{U}_{-(i,t)} = \boldsymbol{u}_{-(i,t)})}{q_{\boldsymbol{\theta}}(U_{i,t} = k, \boldsymbol{U}_{-(i,t)} = \boldsymbol{u}_{-(i,t)})}},$$
which is equivalent to 
\begin{equation}
\log{q_{\boldsymbol{\theta}}(U_{i,t} = w, \boldsymbol{U}_{-(i,t)} = \boldsymbol{u}_{-(i,t)})} - \log{q_{\boldsymbol{\theta}}(U_{i,t} = k, \boldsymbol{U}_{-(i,t)} = \boldsymbol{u}_{-(i,t)})}.
\label{eq:fin}
\end{equation}
Finally, based on Equation \eqref{eq:lat} and \eqref{eq:fin} we have, for $t=1$, that
\small
\begin{equation*}
\begin{split}
& \log{\frac{p(U_{i,1} = w \lvert \boldsymbol{U}_{-(i,1)} = \boldsymbol{u}_{-(i,1)}, \boldsymbol{\theta})}{p(U_{i,1} = k \lvert \boldsymbol{U}_{-(i,1)} = \boldsymbol{u}_{-(i,1)}, \boldsymbol{\theta})}} \\
= & ~ \beta_w -\beta_k + 
\sum_{\substack{j=i+1 \\ j \in \eta_i}}^{N} \left[ \sum_{\substack{v=1 \\ v \neq w}}^K \mathds{1}(U_{i,1}=w, U_{j,1} =v)\gamma_{w,v} - \sum_{\substack{v=1 \\ v \neq k}}^K \mathds{1}(U_{i,1}=k, U_{j,1} =v)\gamma_{k,v} \right] \\
& + \sum_{\substack{v=1 \\ v \neq w}}^K \mathds{1}(U_{i,1} = w, U_{i,2} = v) \delta_{w,v} - \sum_{\substack{v=1 \\v \neq k}}^K \mathds{1}(U_{i,1} = k, U_{i,2} = v) \delta_{k,v},
\end{split}
\end{equation*}
\normalsize
while for $1 < t < T$, we have that
\small
\begin{equation*}
\begin{split}
& \log{\frac{p(U_{i,t} = w \lvert \boldsymbol{U}_{-(i,t)} = \boldsymbol{u}_{-(i,t)}, \boldsymbol{\theta})}{p(U_{i,t} = k \lvert \boldsymbol{U}_{-(i,t)} = \boldsymbol{u}_{-(i,t)}, \boldsymbol{\theta})}} \\
= & ~ \beta^*_w - \beta_k^* +  
\sum_{\substack{j=i+1 \\ j \in \eta_i}}^{N} \left[ \sum_{\substack{v=1 \\ v \neq w}}^K \mathds{1}(U_{i,t}=w, U_{j,t} =v)\gamma^*_{w,v} - \sum_{\substack{v=1 \\ v \neq k}}^K \mathds{1}(U_{i,t}=k, U_{j,t} =v)\gamma^*_{k,v} \right] \\
& + \sum_{\substack{v=1 \\ v \neq w}}^{K} \mathds{1}(U_{i,t} = w, U_{i,t+1} = v) \delta_{w,v} - \sum_{\substack{v=1 \\ v \neq k}}^{K} \mathds{1}(U_{i,t} = k, U_{i,t+1} = v) \delta_{k,v} \\
& + \sum_{\substack{v=1 \\ v \neq w}}^{K} \mathds{1}(U_{i,t-1} = v, U_{i,t} = w) \delta_{v,w} - \sum_{\substack{v=1 \\ v \neq k}}^{K} \mathds{1}(U_{i,t-1} = v, U_{i,t} = k) \delta_{v,k}.
\end{split}
\end{equation*}
\normalsize
Finally, for $t=T$, we have that
\small
\begin{equation*}
\begin{split}
& \log{\frac{p(U_{i,T} = w \lvert \boldsymbol{U}_{-(i,T)} = \boldsymbol{u}_{-(i,T)}, \boldsymbol{\theta})}{p(U_{i,T} = k \lvert \boldsymbol{U}_{-(i,T)} = \boldsymbol{u}_{-(i,T)}, \boldsymbol{\theta})}}  \\
= & ~ \beta^*_w - \beta_k^* + 
\sum_{\substack{j=i+1 \\ j \in \eta_i}}^{N} \left[ \sum_{\substack{v=1 \\ v \neq w}}^K \mathds{1}(U_{i,T}=w, U_{j,T} =v)\gamma^*_{w,v} -  \sum_{\substack{v=1 \\ v \neq k}}^K \mathds{1}(U_{i,T}=k, U_{j,T} =v)\gamma^*_{k,v} \right]  \\
& + \sum_{\substack{v=1 \\ v \neq w}}^{K} \mathds{1}(U_{i,T-1} = v, U_{i,T} = w) \delta_{v,w} - \sum_{\substack{v=1 \\ v \neq k}}^{K}  \mathds{1}(U_{i,T-1} = v, U_{i,T} = k) \delta_{v,k}.
\end{split}
\end{equation*}
\normalsize

\subsection{Multivariate Gaussian response variables and priors}\label{sec:modgaus}

In this work we focus on a (multivariate) Gaussian spatio-temporal HM model, which assumes that

\begin{equation*}\label{eq:3}
\boldsymbol{Y}_{i,t} \lvert U_{i,t} = u \sim \mathcal{N}(\boldsymbol{\mu}_u, \boldsymbol{\Sigma}_u).
\end{equation*}
Obviously, different distributions can be considered instead of the Gaussian one and, with suitable adjustments, it may be possible to include covariates. 

Assuming {\em a priori} independence between parameters, we can write the augmented posterior distribution as 
\begin{equation}\label{eq:posterior}
p(\boldsymbol{\mu}, \boldsymbol{\Sigma}, \boldsymbol{u}, \boldsymbol{\theta} \lvert \boldsymbol{y}) \propto p(\boldsymbol{y} \lvert \boldsymbol{u}, \boldsymbol{\mu}, \boldsymbol{\Sigma}) p(\boldsymbol{u} \lvert \boldsymbol{\theta}) p(\boldsymbol{\mu}) p(\boldsymbol{\Sigma}) p(\boldsymbol{\theta}), 
\end{equation}
where $\boldsymbol{\mu} = \{\boldsymbol{\mu}_1, \ldots, \boldsymbol{\mu}_K\}$, $\boldsymbol{\Sigma} = \{\boldsymbol{\Sigma}_1,\ldots,\boldsymbol{\Sigma}_K\}$, and
$$p(\boldsymbol{y} \lvert \boldsymbol{u}, \boldsymbol{\mu}, \boldsymbol{\Sigma}) \propto \prod_{u = 1}^K \prod_{i =1}^N \prod_{t = 1}^T e^{-\frac{1}{2} (\boldsymbol{y}_{i,t}-\boldsymbol{\mu}_u)'\boldsymbol{\Sigma}_u^{-1}(\boldsymbol{y}_{i,t}-\boldsymbol{\mu}_u) \mathds{1}(U_{i,t}=u)}.$$  
In the previous expression, $p(\boldsymbol{\mu})$ and $p(\boldsymbol{\Sigma})$ are the prior distributions for the vector means and variance-covariance matrices, while $p(\boldsymbol{\theta}) =   p(\boldsymbol{\beta}) p(\boldsymbol{\beta^*})p(\boldsymbol{\gamma})p(\boldsymbol{\gamma^*})p(\boldsymbol{\delta})$ is the product of the prior distributions for the parameters of the latent process. In this setting, the augmented form is also useful when spatio-temporal clustering is considered, since it allows predicting the latent variables using a maximum a posteriori (MAP) approach, instead of decoding methods.

We propose the following prior distributions for the parameters involved in the conditional distribution of the responses 
$$
\boldsymbol{\mu}_u \sim \mathcal{N}(\boldsymbol{m}, \boldsymbol{V}) \quad \text{and} \quad \boldsymbol{\Sigma}_u \sim \mathcal{IW}(\nu, \boldsymbol{S}), 
$$
where $u=1,\ldots,K$. In particular $\boldsymbol{m} \in \mathbb{R}^d$ and $\boldsymbol{V} \in \mathbb{R}^{d \times d}$, while $\mathcal{IW}(\cdot,\cdot)$ denotes
an Inverse-Wishart distribution with degrees of freedom $\nu > d-1$ and a positive definite matrix $\boldsymbol{S} \in \mathbb{R}^{d \times d}$. For the latent distribution parameters, under the assumption of independence, we consider

$$\beta_{u} \sim \mathcal{N}(0, \sigma^2_{\beta_u}) \quad \text{and} \quad \beta^*_{u} \sim \mathcal{N}(0, \sigma^2_{\beta^*_u}), \quad u=1,\ldots,K-1,$$
while
$$\gamma_{u,v} \sim \mathcal{N}(0, \sigma^2_{\gamma_{u,v}}), \quad \gamma^*_{u,v} \sim \mathcal{N}(0, \sigma^2_{\gamma^*_{u,v}}), \quad \text{and} \quad \delta_{u,v} \sim \mathcal{N}(0, \sigma^2_{\delta_{u,v}}), \quad u=1,\ldots,K, ~~ v \neq u.$$

\section{Model estimation}\label{sec:bayes}

We begin this section with a description of the intractable problem when standard MCMC algorithms are considered. Then, we describe the pseudo-posterior solution and finally we introduce the exchange algorithm and its approximate version. 

\subsection{Bayesian inference}\label{sec:inf}

It is easy to verify that a classical MCMC algorithm for the parameters of the latent variables
cannot be used in Equation \eqref{eq:posterior} since the normalizing constant $\mathcal{Z}_{\boldsymbol{\theta}}$ depends on $\boldsymbol{\theta}$ and does not simplify in the acceptance rate. For example, consider the following naïve MH algorithm with symmetric proposal and acceptance probability 

\begin{equation}\label{eq:ac}
\alpha(\boldsymbol{\theta}, \boldsymbol{\tilde{\theta}})= 1 \wedge \frac{p(\boldsymbol{\tilde{\theta}}) p(\boldsymbol{u} \lvert \boldsymbol{\tilde{\theta}})}{p(\boldsymbol{\theta}) p(\boldsymbol{u} \lvert \boldsymbol{\theta})} = 1 \wedge \frac{p(\boldsymbol{\tilde{\theta}}) q_{\boldsymbol{\tilde{\theta}}}(\boldsymbol{u})}{p(\boldsymbol{\theta}) q_{\boldsymbol{\theta}}(\boldsymbol{u})} \frac{\mathcal{Z}_{\boldsymbol{\theta}}}{\mathcal{Z}_{\boldsymbol{\tilde{\theta}}}},
\end{equation}
where $p(\boldsymbol{u} \lvert \boldsymbol{\theta}) = q_{\boldsymbol{\theta}} (\boldsymbol{u})/\mathcal{Z}_{\boldsymbol{\theta}}$.
The normalizing constants in Equation \eqref{eq:ac} are computationally intractable, and since $\mathcal{Z}_{\boldsymbol{\tilde{\theta}}} \neq \mathcal{Z}_{\boldsymbol{\theta}}$ they do not simplify. A possible solution is based on the pseudo-posterior distribution, where the distribution $p(\boldsymbol{u} \lvert \boldsymbol{\theta})$ is replaced by a pseudo-distribution; see \cite{bouranis2017efficient}. This pseudo-distribution is typically defined as follows

\begin{equation*}\label{pseudo}
p_{\text{pseudo}}(\boldsymbol{u} \lvert 
\boldsymbol{\theta}) = \prod_{i=1}^{N} \left[ p(u_{i,1} \lvert \boldsymbol{u}_{-(i,1)}, \boldsymbol{\theta}) \prod_{t>1} p(u_{i,t} \lvert \boldsymbol{u}_{-(i,t)}, \boldsymbol{\theta}) \right],
\end{equation*}
where a product of full conditionals is involved. Using this distribution, the MCMC target becomes

 \begin{equation*}\label{eq:pseudoposterior}
p_{\text{pseudo}}(\boldsymbol{\mu}, \boldsymbol{\Sigma}, \boldsymbol{u}, \boldsymbol{\theta} \lvert \boldsymbol{y}) \propto p(\boldsymbol{y} \lvert \boldsymbol{u}, \boldsymbol{\mu}, \boldsymbol{\Sigma}) p_{\text{pseudo}}(\boldsymbol{u} \lvert \boldsymbol{\theta}) p(\boldsymbol{\mu}) p(\boldsymbol{\Sigma}) p(\boldsymbol{\theta}), 
\end{equation*}
which is now tractable. The acceptance probability in Equation \eqref{eq:ac} becomes

\begin{equation*}\label{eq:ps}
\alpha(\boldsymbol{\theta}, \boldsymbol{\tilde{\theta}})= 1 \wedge \frac{p(\boldsymbol{\tilde{\theta}}) p_{\text{pseudo}}(\boldsymbol{u} \lvert 
\boldsymbol{\tilde{\theta}})}{p(\boldsymbol{\theta}) p_{\text{pseudo}}(\boldsymbol{u} \lvert 
\boldsymbol{\theta})},
\end{equation*}
where the ratio of normalizing constants is not involved. Notice that, when possible, a calibration of the pseudo-posterior distribution can be obtained as in \cite{bouranis2017efficient}.

An alternative solution to the pseudo-distribution approach is the algorithm proposed by \cite{moller2006efficient}, while its extension is the widely used exchange algorithm of \cite{murray2012mcmc}.
In our case, the exchange algorithm samples from a posterior distribution which is augmented by an auxiliary process 
$$\{\Omega_{i,t}\} = \{ \Omega_{i,t} : (i,t) \in \mathcal{S} \times \mathcal{T}\},$$
which has to be $\{\Omega_{i,t}\} \stackrel{d}{=} \{U_{i,t}\}$, meaning that the auxiliary process has to be equal in distribution to the latent one, and it should be generated by a perfect sampler.  
In our setting, instead of using Equation \eqref{eq:posterior}, we consider the following target
\begin{equation}\label{eq:posterior_exchange}
p(\boldsymbol{\mu}, \boldsymbol{\Sigma}, \boldsymbol{u}, \boldsymbol{\omega}, \boldsymbol{\theta},  \boldsymbol{\tilde{\theta}} \lvert \boldsymbol{y}) \propto p(\boldsymbol{y} \lvert \boldsymbol{u}, \boldsymbol{\mu}, \boldsymbol{\Sigma}) p(\boldsymbol{u} \lvert \boldsymbol{\theta}) p(\boldsymbol{\mu}) p(\boldsymbol{\Sigma}) p(\boldsymbol{\theta}) h(\boldsymbol{\tilde{\theta}} \lvert \boldsymbol{\theta}) p(\boldsymbol{\omega} \lvert \boldsymbol{\tilde{\theta}}), 
\end{equation}
where $h(\boldsymbol{\tilde{\theta}} \lvert \boldsymbol{\theta})$ is typically a symmetric density distribution and $\boldsymbol{\omega} = \{\omega_{1,1},\ldots,\omega_{N,T}\}$ denotes the realized collection of auxiliary variables. Notice that, in Equation \eqref{eq:posterior_exchange}, if $\boldsymbol{\omega}$ and $\tilde{\boldsymbol{\theta}}$ are integrated out then the posterior distribution in Equation \eqref{eq:posterior} is obtained, which justifies the use of the augmented distribution. 

Assuming that a perfect sampler for $\boldsymbol{\omega}$ exists, the original exchange algorithm is outlined as follows:

\begin{enumerate}
\item draw $\boldsymbol{\tilde{\theta}} \sim h(\cdot \lvert \boldsymbol{\theta})$;
\item draw $\boldsymbol{\omega} \sim p(\cdot \lvert \boldsymbol{\tilde{\theta}})$;
\item accept the swap $\boldsymbol{\theta}$ to $\boldsymbol{\tilde{\theta}}$ with probability \[1 \wedge \frac{q_{\boldsymbol{\tilde{\theta}}}(\boldsymbol{u}) p(\boldsymbol{\tilde{\theta}}) h(\boldsymbol{\theta} \lvert \boldsymbol{\tilde{\theta}}) q_{\boldsymbol{\theta}}(\boldsymbol{\omega})}{q_{\boldsymbol{\theta}}(\boldsymbol{u}) p(\boldsymbol{\theta}) h(\boldsymbol{\tilde{\theta}} \lvert \boldsymbol{\theta}) q_{\boldsymbol{\tilde{\theta}}}(\boldsymbol{\omega})}  \frac{\mathcal{Z}_{\boldsymbol{\theta}} \mathcal{Z}_{\boldsymbol{\tilde{\theta}}}}{\mathcal{Z}_{\boldsymbol{\tilde{\theta}}} \mathcal{Z}_{\boldsymbol{\theta}}}.\]
\end{enumerate}
In the last step, the ratio of normalizing constants simplifies to 1, allowing us to evaluate the acceptance probability of the Markov chain. We briefly discuss the ratio obtained in the previous point. There is a clear relation between $q_{\boldsymbol{\theta}}(\boldsymbol{\omega})/q_{\tilde{\boldsymbol{\theta}}}(\boldsymbol{\omega})$ and $\mathcal{Z}_{\boldsymbol{\theta}}/\mathcal{Z}_{\tilde{\boldsymbol{\theta}}}$;
in fact, we have that
$$\mathbb{E}_{p(\boldsymbol{\omega}  \lvert \tilde{\boldsymbol{\theta}})} \left[\frac{q_{\boldsymbol{\theta}}(\boldsymbol{\omega})}{q_{\tilde{\boldsymbol{\theta}}}(\boldsymbol{\omega})}\right] = \frac{\mathcal{Z}_{\boldsymbol{\theta}}}{\mathcal{Z}_{\tilde{\boldsymbol{\theta}}}},$$
and so one could consider the following approximation
$$\frac{q_{\boldsymbol{\tilde{\theta}}}(\boldsymbol{u}) p(\boldsymbol{\tilde{\theta}}) h(\boldsymbol{\theta} \lvert \boldsymbol{\tilde{\theta}}) q_{\boldsymbol{\theta}}(\boldsymbol{\omega})}{q_{\boldsymbol{\theta}}(\boldsymbol{u}) p(\boldsymbol{\theta}) h(\boldsymbol{\tilde{\theta}} \lvert \boldsymbol{\theta}) q_{\boldsymbol{\tilde{\theta}}}(\boldsymbol{\omega})} \approx \frac{q_{\boldsymbol{\tilde{\theta}}}(\boldsymbol{u}) p(\boldsymbol{\tilde{\theta}}) h(\boldsymbol{\theta} \lvert \boldsymbol{\tilde{\theta}}) \mathcal{Z}_{\boldsymbol{\theta}}}{q_{\boldsymbol{\theta}}(\boldsymbol{u}) p(\boldsymbol{\theta}) h(\boldsymbol{\tilde{\theta}} \lvert \boldsymbol{\theta}) \mathcal{Z}_{\tilde{\boldsymbol{\theta}}}}.$$
We can consider an unbiased estimator of $\mathcal{Z}_{\boldsymbol{\theta}}/\mathcal{Z}_{\tilde{\boldsymbol{\theta}}}$ obtained as follows
$$\frac{1}{J} \sum_{j=1}^J \frac{q_{\boldsymbol{\theta}}(\boldsymbol{\omega}_j)}{q_{\tilde{\boldsymbol{\theta}}}(\boldsymbol{\omega}_j)} \approx \frac{\mathcal{Z}_{\boldsymbol{\theta}}}{\mathcal{Z}_{\tilde{\boldsymbol{\theta}}}},$$
the resulting algorithm has been labeled noisy exchange algorithm \citep{alquier2016noisy}. Notice that when $J = 1$ the framework corresponds to the exchange algorithm, whereas when $J \rightarrow \infty$ the standard MH algorithm ensues. Setting $J = 1$ or $J \rightarrow \infty$ leaves the target posterior invariant. Unfortunately, when $1 < J < \infty$ the noisy exchange is not guaranteed to sample from the posterior; see \cite{alquier2016noisy} for details.

In our setting, a perfect sampler for $\boldsymbol{\omega}$ is not available; however, an alternative MCMC called approximate exchange algorithm has been proposed in \cite{friel2011classification}, where the exact auxiliary sampler is substituted by a Gibbs sampler. In particular, the auxiliary process is obtained from the last iteration of a Gibbs sampler. Theoretical justifications, based on mild assumptions, for using the final iteration can be found in \cite{everitt2012bayesian}. In particular, when the MCMC kernel for the exact exchange algorithm is uniformly ergodic, the invariant distribution of the corresponding approximate exchange algorithm becomes closer to the “true” target (that of the exact exchange algorithm) as the number of the auxiliary Gibbs iterations increases. For more details, we refer the reader to the supplementary material in \cite{everitt2012bayesian}, specifically Theorem 2 in Appendix B.

In order to estimate the model proposed, we consider an MCMC algorithm combining the approximate exchange and Gibbs steps. From Equation (\ref{eq:posterior_exchange}), we can obtain the following full conditional distributions; full calculations are provided in Appendix B. For the mean vectors we have
\begin{equation}\label{eq:mean}
\boldsymbol{\mu}_u \lvert \cdots \sim \mathcal{N}( \tilde{\boldsymbol{V}}_u \tilde{\boldsymbol{m}}_u, \tilde{\boldsymbol{V}}_u ), 
\end{equation}  
where
$$
\tilde{\boldsymbol{V}}_u^{-1} = n_u \boldsymbol{\Sigma}_u^{-1}+ \boldsymbol{V}^{-1} ~~~ \text{and} ~~~ \tilde{\boldsymbol{m}}_u =  \boldsymbol{\Sigma}_u^{-1} n_u{\boldsymbol{\bar{y}}_u} + \boldsymbol{V}^{-1}\boldsymbol{m},
$$
with 
$$n_u = \sum_{i=1}^{N} \sum_{t = 1}^T \mathds{1}(U_{i,t}=u) ~~~ \text{and}~~~ {\boldsymbol{\bar{y}}_u} = (1/n_u) \sum_{i=1}^N \sum_{t=1}^{T} \boldsymbol{y}_{i,t} \mathds{1}(U_{i,t}=u).$$ 
For the variance-covariance 
matrices we have
\begin{equation}\label{eq:variance}
\boldsymbol{\Sigma}_u \lvert \cdots \sim \mathcal{IW}(\nu + n_u, \boldsymbol{S} + \tilde{\boldsymbol{S}}_u), 
\end{equation}
where
$$
\tilde{\boldsymbol{S}}_u = \sum_{i=1}^N \sum_{t=1}^{T} (\boldsymbol{y}_{i,t} -\boldsymbol{\mu}_u)(\boldsymbol{y}_{i,t}-\boldsymbol{\mu}_u)' \mathds{1}(U_{i,t}=u).
$$ 
These full conditionals can be obtained following the same approach as 
in \cite{tancini2024comparison}. In the approximate exchange steps, we update each parameter in $\boldsymbol{\theta}$ using an individual move. For each $\beta_u$ we propose a move consisting in generating a new $\tilde{\beta}_u$ using a random walk
$$
\tilde{\beta}_u = \beta_u + \epsilon_{\beta_u}, 
$$
where $\epsilon_{\beta_u} \sim \mathcal{N}(0, \phi^2_{\beta_u})$, and then accepting the swap $\beta_u$ to 
$\tilde{\beta}_u$ with probability
\begin{equation*}\label{eq:b}
1 \wedge \frac{p(\tilde{\beta}_u) q(\boldsymbol{u}; \boldsymbol{\tilde{\beta}}, \boldsymbol{\beta^*}, \boldsymbol{\delta}, \boldsymbol{\gamma}, \boldsymbol{\gamma^*}) q(\boldsymbol{\omega}; \boldsymbol{\beta}, \boldsymbol{\beta^*}, \boldsymbol{\delta}, \boldsymbol{\gamma}, \boldsymbol{\gamma^*})}{p(\beta_u) q(\boldsymbol{u}; \boldsymbol{\beta}, \boldsymbol{\beta^*}, \boldsymbol{\delta}, \boldsymbol{\gamma}, \boldsymbol{\gamma^*}) q(\boldsymbol{\omega}; \boldsymbol{\tilde{\beta}}, \boldsymbol{\beta^*}, \boldsymbol{\delta}, \boldsymbol{\gamma}, \boldsymbol{\gamma^*})},
\end{equation*}
where $q(\boldsymbol{u}; \boldsymbol{\tilde{\beta}}, \boldsymbol{\beta^*}, \boldsymbol{\delta}, \boldsymbol{\gamma}, \boldsymbol{\gamma^*}) = q_{\boldsymbol{\tilde{\theta}}}(\boldsymbol{u})$ and $\boldsymbol{\omega} \sim q_{\boldsymbol{\tilde{\theta}}}(\cdot)/\mathcal{Z}_{\boldsymbol{\tilde{\theta}}}$, taking the last iteration of a Gibbs sampler for $\boldsymbol{\omega}$. 

A critical point is to define the number of auxiliary iterations required for the Gibbs sampler. This point has been analyzed in \cite{bhamidi2008mixing}, where it is shown that convergence of sampling from a large scale ERGM framework through MCMC is likely to be slow. In addition, the same authors suggest to take a conservative approach and choose a large number of auxiliary iterations. To manage this computational problem, we provide a specific initialization strategy in Section \ref{sec:in} which tries to reduce the number of auxiliary iterations.

An update step analogous to that of $\beta_u$ is followed for each parameter in $\boldsymbol{\beta^*}$, $\boldsymbol{\gamma}$, $\boldsymbol{\gamma^*}$, and $\boldsymbol{\delta}$. For the approximate exchange steps we consider an adaptive vanishing procedure. In particular, we consider the global adaptive scaling described in \citet[Section 5.1.2]{andrieu2008tutorial}, where the mean and the variance of the random walk are updated at each iteration of the algorithm using a scale parameter. The scale parameter is adapted according to three components: the desirable acceptance probability, the acceptance probability evaluated at each iteration, and a decreasing stepsize sequence.
We set the stepsize sequence as $C/r$, where $C \in \mathbb{R}_+$ and $r \in \{1,\ldots,R\}$ is the iteration counter of the algorithm, with $R$ being the maximum number of iterations. We consider this approach for the first $50\%$ of the chain, setting then the stepsize to 0. 

Finally, for the latent variables we use the following full conditionals:

\vspace{0.3cm}

\begin{equation}
\begin{split}
p(U_{i,1} = u \lvert \ldots) \propto & ~p(\boldsymbol{y}_{i,1} \lvert U_{i,1} = u, \boldsymbol{\mu}_u, \boldsymbol{\Sigma}_u) ~ \text{exp}\bigg\{ \beta_u + \delta_{u,u_{i,t+1}}  \\
& + \sum_{\substack{j=i+1 \\ j \in \eta_i}}^{N} \mathds{1}(U_{i,1} = u, U_{j,1} = u_{j,1}) \gamma_{u,u_{j,1}} \bigg\},
\end{split}
\label{eq:t}
\end{equation}

for $t=1$;

\begin{equation}
\begin{split}
p(U_{i,t} = u \lvert \ldots) \propto &~ p(\boldsymbol{y}_{i,t} \lvert U_{i,t} = u, \boldsymbol{\mu}_u, \boldsymbol{\Sigma}_u) ~ \text{exp}\bigg\{ \beta^*_u + \delta_{u_{i,t-1},u} + \delta_{u,u_{i,t+1}}  \\
& + \sum_{\substack{j=i+1 \\ j \in \eta_i}}^{N} \mathds{1}(U_{i,t} = u, U_{j,t} = u_{j,t}) \gamma^*_{u,u_{j,t}} \bigg\},
\end{split}
\label{eq:Tt}
\end{equation}

for $1 < t < T$;

\begin{equation}
\begin{split}
p(U_{i,T} = u \lvert \ldots) \propto & ~p(\boldsymbol{y}_{i,T} \lvert U_{i,T} = u, \boldsymbol{\mu}_u, \boldsymbol{\Sigma}_u) ~ \text{exp} \bigg\{ \beta^*_u + \delta_{u_{i,T-1},u}  \\
& + \sum_{\substack{j=i+1 \\ j \in \eta_i}}^{N} \mathds{1}(U_{i,T} = u, U_{j,T} = u_{j,T}) \gamma^*_{u,u_{j,T}} \bigg\}.
\end{split}
\label{eq:T}
\end{equation}

for $t=T$.

\vspace{0.3cm}

Clearly, we need a Gibbs sampler for the auxiliary process $\{\Omega_{i,t}\}$. It is easy to prove that the full conditional distributions required for $\{\Omega_{i,t}\}$ are equal to those reported in Equation \eqref{eq:t}, \eqref{eq:Tt}, and \eqref{eq:T}, once the $p(\boldsymbol{y}_{i,t} \lvert U_{i,t} = u, \boldsymbol{\mu}_u, \boldsymbol{\Sigma}_u)$ part is removed.

\subsection{Auxiliary variable initialization}\label{sec:in}

The problem of the number of auxiliary iterations of the Gibbs sampler used in the approximate exchange algorithm has been studied in different works. \cite{caimo2011bayesian} suggested that 500 iterations is a long-enough run for ERGM, while \cite{everitt2012bayesian} suggested that 50 to 100 iterations are usually sufficient when latent Markov random fields are considered. \cite{bhamidi2008mixing} showed that MCMC-based sampling for large ERGM often suffers from exponentially slow convergence. To address this issue, they advocate for a conservative approach with many auxiliary iterations; however, this renders the exchange algorithm computationally infeasible for large graphs. Since we include a latent process with similar structures to those previously defined, including both spatial and temporal dependence in the model, we expect that the same computational problem described above arises.

In this section, we consider a possible approach which tries to decrease the number of iterations required for the auxiliary process, leading to a significant reduction of computational time of the algorithm. For each parameter in the exchange steps, we generate an auxiliary process from a Gibbs sampler, considering a fixed number of iterations $M$. The auxiliary $\boldsymbol{\omega}$ lives in the same space as $\boldsymbol{u}$, which is latent in the model, in particular 
$\boldsymbol{\omega} \sim p(\cdot \lvert \boldsymbol{\theta})$. 

The idea is to initialize the Gibbs for the auxiliary $\boldsymbol{\omega}$ by taking the most recent value of $\boldsymbol{u}$.
We expect, heuristically, that the initial region of the latent process from the previous iteration should be closer to the auxiliary process, as each parameter in $\boldsymbol{\theta}$ is updated individually. Using this initialization we can dramatically reduce the number of auxiliary iterations. 

The previous approach can also be further refined by including a non-increasing function for the number of auxiliary iterations required for the Gibbs sampler. This means that 
we can initially take a more conservative approach, whereby for the first few iterations we use a higher number of auxiliary iterations. However, as the number of iterations increases and as the chain moves to a better area of the sample space, we can potentially reduce the number of auxiliary iterations.

\section{Simulation study}\label{sec:sim}

In this section, we present a simulation study designed to evaluate the performance of the approximate exchange algorithm and the pseudo-posterior MCMC algorithm, as described in Section \ref{sec:bayes}. We consider four distinct scenarios, in detail Scenarios A, B, C, and D, each are outlined in detail in the following.
We generate 50 simulated datasets under each 
scenario 
and we compute the mean absolute error (MAE) of the estimated parameters, averaging over the 50 samples. This allows us to provide a robust summary of the performances of the two methods under different data-generating conditions.

In addition, we focus on two representative synthetic datasets, selected from Scenarios A and C. These datasets are used for illustrative purposes to provide a more in-depth understanding of the behavior of the two methods. For each of these datasets, we compare the posterior distributions obtained from both algorithms, reporting the posterior expectations and the corresponding Monte Carlo standard errors \citep{flegal2010batch}. We also assess the convergence of the MCMC chains using standard diagnostic tools, and we evaluate the classification accuracy through the misclassification rate.

All algorithms have been implemented in R, and the source code is available at the following link: 
\url{https://github.com/DanieleTancini/Spatio-temporal-HMM}.

\subsection{Simulation design}\label{sec:simdes}

We consider a simulation study to compare the approximate exchange algorithm and the pseudo-posterior MCMC. First, notice that the neighbourhood system, for all $t \in \mathcal{T}$, can be defined using a graph $\mathcal{G}=(\mathcal{W},\mathcal{E})$ with $\mathcal{W}$ being the set of nodes and $\mathcal{E}$ that of edges. This means that neighbourhood systems can be randomly generated using network models. 

In this simulation study, we consider 4 different scenarios, and for each of them, we generate $D = 50$ random datasets. Due to the nature of the data typically used with spatio-temporal HM models, where sites represent regions or countries over relatively short time periods, typically measured by years, we do not consider the simulated data to be high-dimensional. On the contrary, since the spatial dependence plays a central role in these type of models, we analyze different spatial structures.

\begin{figure}[h]
\centering
\includegraphics[width=0.3\textwidth]{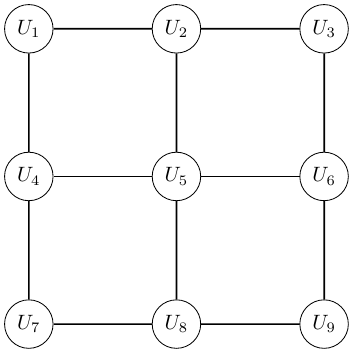} 
\caption{Regular neighbourhood system.} 
\label{fig0}
\end{figure}

For Scenario A we fix $N=9$ and $T=5$, while for the number of states we use $K=2$ and a regular neighbourhood system (such as a regular square grid) is used. Let $z \times z 
$ be the dimension of the square grid, constructed with $2z(z-1)$ edges. A graphical representation of the regular square grid is reported in Figure \ref{fig0}.

For the set of parameters of the observable variables, which includes $\boldsymbol{\mu}$ and $\boldsymbol{\Sigma}$, we set
$$\boldsymbol{\mu}_1 = (-3,-3)', \quad \boldsymbol{\mu}_2 = (3,3)', \quad \text{and} \quad 
\boldsymbol{\Sigma}_1 = \boldsymbol{\Sigma}_2 = \begin{pmatrix}
1 & 0 \\
0 & 1 
\end{pmatrix}.$$
These parameters are not of main interest in this simulation study, as both the approximate exchange algorithm and pseudo-posterior MCMC have closed-form solutions for the full conditional distributions, meaning that the same Gibbs steps can be used. We do not expect significant differences in their estimations, and for this reason we keep them unchanged across Scenarios A, B, and C. For the latent process, we set the following parameters:

$$\beta_1 = 2, \quad \beta^*_1 = 2, \quad \boldsymbol{\gamma} = \boldsymbol{\gamma^*} = \begin{pmatrix}
0 & -1 \\
1 & 0 
\end{pmatrix}, \quad \text{and} \quad \boldsymbol{\delta} = \begin{pmatrix}
0 & -1 \\
-1 & 0 
\end{pmatrix}.$$
We use these parameters in order to obtain realistic synthetic datasets, avoiding models which include only one state, or where there are empty classes. In particular, we consider a framework where nodes are more likely to persist in the same class rather than moving into a different one. This is obtained considering positive values for $\beta_1$ and $\beta^*_1$, including negative values for $\gamma_{1,2}$ and $\gamma^*_{1,2}$, and negative values for $\delta_{1,2}$ and $\delta_{2,1}$. Notice that, in this scenario, the neighbourhood systems is equal in each sample $D$, due to the regularity of the square grid lattice. 

In Scenario B, the number of sites is increased to $N=40$, maintaining $T=5$, and $K=2$. 
For the set of parameters of the observable variables we use the same parameters proposed in the previous scenario. The neighbourhood system is generated using the Erdős-Renyi model \citep{erdds1959random}, choosing a graph uniformly at random from the collection of all graphs which have 
40 nodes and 20 edges. 
Since the spatial structure is randomly sampled, different patterns of dependencies may arise. This approach allows us to explore various types of spatial structures rather than repeatedly analyzing the same structure 
$D$ times. 
For the latent process, we set the following parameters:

$$\beta_1 = 2, \quad \beta^*_1 = 2, \quad \boldsymbol{\gamma} = \boldsymbol{\gamma^*} = \begin{pmatrix}
0 & -2 \\
2 & 0 
\end{pmatrix}, \quad \text{and} \quad \boldsymbol{\delta} = \begin{pmatrix}
0 & -2 \\
-2 & 0 
\end{pmatrix}.$$
This configuration is chosen for the same reasons explained for the previous scenario.

In Scenario C, the number of times is increased to $T=10$, maintaining the same $N$ and $K$. The neighbourhood system is still generated using the Erdős-Renyi model. The random samples are generated using this set of parameters for the latent process: 

$$\beta_1 = 2, \quad \beta^*_1 = 2, \quad \boldsymbol{\gamma} = \boldsymbol{\gamma^*} = \begin{pmatrix}
0 & -2 \\
2 & 0 
\end{pmatrix}, \quad \text{and} \quad \boldsymbol{\delta} = \begin{pmatrix}
0 & -1 \\
-1 & 0 
\end{pmatrix}.$$
The aim of this simulation study is to investigate the behavior of the pseudo-posterior approach and the approximate exchange algorithm when time increase, maintaining the same spatial structure used in Scenario B.

Finally, a fourth scenario is analyzed, denoted as Scenario D. In this scenario, we consider the same setting analyzed for Scenario B, that is, $N=40$ and $T=5$, but we increase the number of states to $K=3$ generating $D = 50$ samples. The neighbourhood system is still defined using the Erdős-Renyi model. In this scenario, the following parameter values are used:

$$\beta_1 = \beta_2 = 0,  \quad \text{and} \quad \beta^*_1 = \beta^*_2 = 0,$$
while 
$$ \boldsymbol{\gamma} = \boldsymbol{\gamma^*} = \begin{pmatrix}
0 & -2 & -2\\
-2 & 0 &-2 \\
-2 & -2 & 0
\end{pmatrix}, \quad \text{and} \quad \boldsymbol{\delta} = \begin{pmatrix}
0 & -1 & -1\\
-1 & 0 &-1 \\
-1 & -1 & 0
\end{pmatrix}.$$
Since we consider $K=3$, we impose a different setting of $\boldsymbol{\mu}$ and $\boldsymbol{\Sigma}$, and, in particular, we use:
$$\boldsymbol{\mu}_1 = (-5,-5)', \quad \boldsymbol{\mu}_2 = (0,5)', \quad \boldsymbol{\mu}_2 = (5,-5)', \quad \text{and} \quad \boldsymbol{\Sigma}_1 = \boldsymbol{\Sigma}_2 = \boldsymbol{\Sigma}_3 = \begin{pmatrix}
1 & 0 \\
0 & 1 
\end{pmatrix}.$$
Also in this case, the $\boldsymbol{\mu}_u$ and $\boldsymbol{\Sigma}_u$ are not the primary focus of this simulation study, as both the approximate exchange algorithm and pseudo-posterior MCMC can efficiently estimate these parameters using full conditionals in standard form.

For a fair comparison, the same starting values are used for both competing algorithms including the latent process, where each latent variable is sampled from a categorical distribution with uniform probabilities $1/K$.
Each algorithm is run for 10,000 iterations, discarding the first 5,000 samples as burn-in. We use the following hyperparameters for the priors introduced in Section \ref{sec:model}: 

\begin{itemize}
\item $\boldsymbol{m} = \boldsymbol{0}$ and $\boldsymbol{V} = 100 \boldsymbol{I}$, where $\boldsymbol{I}$ is an identity matrix, for all $u=1,\ldots,K$;
\item $\nu = 2\{\text{int}[(d+1)/2] + 1\}$ and $\boldsymbol{S} = (s_{h,l})$ with $h,l=1,\ldots,d$ such that

\[ 
s_{h,l} = \begin{cases} 
  \nu  & \quad \text{if} \quad h=l \\ 
  \pm \nu/2 &\quad \text{if} \quad h\neq l
\end{cases}
\]

for all $u=1,\ldots,K$, where $\text{int}(\cdot)$ is the greatest integer function, and $d$ is the dimension of the response variable $\boldsymbol{Y}_{i,t} \in \mathbb{R}^d$, obtaining minimal informative priors on the variance covariance matrix as in \cite{spezia2010bayesian};

\item $\sigma_{\beta_u}^2 = \sigma_{\beta^*_u}^2 = \sigma_{\gamma_{u,v}}^2 = \sigma_{\gamma^*_{u,v}}^2 = \sigma_{\delta_{u,v}}^2 = 1$, for $u=1,\ldots,K$, $v \neq u$.

\end{itemize}

We briefly discuss the last point, related to the adopted $\sigma_{\beta_u}^2,\ldots,\sigma^2_{\delta_{u,v}}$. Since we generate multiple data for each scenario, considering different spatial structures (randomly sampled), we impose small-variance hyperparameters for the parameters associated to the latent process for both the approximate exchange and the pseudo-posterior approach, with the aim to mitigate possible issues related to the generation of empty latent classes, which can be common in these types of model.

For the auxiliary variable required in the approximate exchange algorithm, we consider the initialization strategy proposed in Section \ref{sec:in}, and we use only five iterations for each Gibbs sampler associated to the auxiliary variable. This number has been obtained considering different trials, starting from a single auxiliary iteration, evaluating then the samples of the approximate exchange algorithm. Notice that this approach reduces the computational time required for the approximate exchange algorithm, in contrast to the larger values suggested in \cite{everitt2012bayesian} and \cite{caimo2011bayesian}.

\subsection{Simulations results for Scenarios A, B, C, and D}

Our analysis begins with the evaluation of all simulated datasets across various scenarios. For this broader comparison, we use the MAE of the estimated parameters as a summary measure of estimation accuracy, providing a comprehensive view of performance across different data-generating conditions. 

Finally, we conclude our evaluation with a detailed examination of results from two synthetic datasets. For each method, we evaluate the quality of the posterior samples by comparing the posterior expectations and the corresponding Monte Carlo standard errors, following the approach described in \citet{flegal2010batch}. We also assess convergence and sampling quality using standard diagnostic tools. In addition, we assess the classification performance of each method by computing the misclassification rate. This provides insight into how well each approach can recover the latent structure in the data.

We begin our discussion with the results from Scenario A, in which the spatial structure remains regular and consistent across all 50 generated datasets. The MAE for each parameter of the latent process associated to Scenario A is reported in Table \ref{tab4}.
\begin{table}[h!]
\begin{center}
\caption{MAE of the approximate exchange and pseudo-posterior algorithms computed for each parameter across 50 samples, evaluated under Scenario A. The lowest values between the approximate exchange and the pseudo-posterior algorithm are reported in bold for each parameter.} 
\vspace{0.1cm}
{\begin{tabular}{ccc}
\hline
& \multicolumn{2}{c}{Mean absolute error} \\
 Parameter & Approx. exchange & Pseudo-post. \\ 
\hline 
 $\beta_{1}$ & \textbf{1.074} & 1.472   \\ 
 $\beta^*_{1}$&  \textbf{0.401} & 1.634 \\ 
  $\gamma_{1,2} $ &  \textbf{0.674} & 0.908 \\ 
  $\gamma_{2,1}$&  \textbf{0.738} & 1.119  \\ 
   $\gamma^*_{1,2} $ &  \textbf{0.871} & 0.936 \\ 
   $\gamma^*_{2,1}$&  \textbf{0.617} & 1.249  \\ 
  $\delta_{1,2} $ &  \textbf{0.338} & 0.561 \\ 
   $\delta_{2,1}$& 0.437 &  \textbf{0.412} \\ 
\hline
\end{tabular}}\label{tab4}
\end{center}
\end{table}

The estimates obtained using the approximate exchange algorithm are generally more accurate than those produced by the pseudo-posterior method. This implies lower MAEs across all parameters, except for of $\delta_{2,1}$, where both methods yield similar results.  

In Scenario B, the spatial structure is different for each generated dataset, since it is randomly obtained from an Erdős-Renyi model. These results are reported in Table \ref{tab5}. 
\begin{table}[h!]
\begin{center}
\caption{MAE of the approximate exchange and pseudo-posterior algorithms computed for each parameter across 50 samples, evaluated under Scenario B. The lowest values between the approximate exchange and the pseudo-posterior algorithm are reported in bold for each parameter.} 
\vspace{0.1cm}
{\begin{tabular}{ccc}
\hline
 & \multicolumn{2}{c}{Mean absolute error} \\
 Parameter & Approx. exchange & Pseudo-post. \\ 
\hline 
 $\beta_{1}$ &  \textbf{1.010} & 1.102 \\ 
 $\beta^*_{1}$&  \textbf{0.535} & 1.458 \\ 
 $\gamma_{1,2}$&  \textbf{1.550} & 1.860 \\ 
  $\gamma_{2,1} $ &  \textbf{0.760} & 1.315 \\ 
   $\gamma^*_{1,2}$&  \textbf{1.091} & 1.856  \\ 
  $\gamma^*_{2,1} $ &  \textbf{0.593} & 1.486 \\ 
  $\delta_{1,2}$&  \textbf{0.338} & 0.701  \\ 
  $\delta_{2,1} $ & 0.418 &  \textbf{0.345} \\ 
 \hline 
\end{tabular}}\label{tab5}
\end{center}
\end{table}

As in the first scenario, the values obtained using the approximate exchange algorithm are more precise than those produced by the pseudo-posterior algorithm, except for $\delta_{2,1}$. These results suggest that the spatial structure inherent to the model does not affect the comparison in terms of performances of the two algorithms.

Looking at Scenario C, the spatial structure is still random and generated from an Erdős-Renyi model, but the number of observations in the datasets is larger since the number of times is increased from $T=5$ to $T=10$. The MAEs obtained are reported in Table \ref{tab6}. 
\begin{table}[h!]
\begin{center}
\caption{MAE of the approximate exchange and pseudo-posterior algorithms computed for each parameter across 50 samples, evaluated under Scenario C. The lowest values between the approximate exchange and the pseudo-posterior algorithm are reported in bold for each parameter.} 
\vspace{0.1cm}
{\begin{tabular}{ccc}
\hline
 & \multicolumn{2}{c}{Mean absolute error} \\
  Parameter & Approx. exchange & Pseudo-post. \\ 
\hline 
  $\beta_{1}$ &  \textbf{0.845} & 0.952 \\ 
 $\beta^*_{1}$&  \textbf{0.480} & 1.420 \\ 
 $\gamma_{1,2}$&  \textbf{1.475} & 1.760 \\ 
  $\gamma_{2,1} $ &  \textbf{0.573} & 1.093 \\ 
   $\gamma^*_{1,2}$&  \textbf{0.634} & 1.775  \\ 
  $\gamma^*_{2,1} $ &  \textbf{0.516} & 1.442 \\ 
  $\delta_{1,2}$&  \textbf{0.473} & 1.338 \\ 
  $\delta_{2,1} $ &  \textbf{0.693} & 0.772 \\ 
\hline 
\end{tabular}}\label{tab6}
\end{center}
\end{table}

Also in Scenario C, the approximate exchange algorithm yields more accurate estimates than the pseudo-posterior method, as evidenced by consistently lower MAEs over all parameters. In addition, comparing the results obtained from Scenarios B and C, the MAEs values obtained in Scenario C are lower than those obtained in Scenario B, as expected, since the number of observations increases. These results show better performance of the approximate exchange algorithm even when we modify the number of time points.

Finally, we discuss the results obtained in Scenario D, where we increase $K$ from 2 to 3, taking $N=40$ and $T=5$. As for the previous case, the spatial structure varies across generated datasets, as it is randomly generated from an Erdős-Rényi model. The key difference in this scenario compared to the others analyzed is the increased number of parameters. The results obtained are reported in Table \ref{tab7}.

\begin{table}[h!]
\begin{center}
\caption{MAE of the approximate exchange and pseudo-posterior algorithms computed for each parameter across 50 samples, evaluated under Scenario D. The lowest values between the approximate exchange and the pseudo-posterior algorithm are reported in bold for each parameter.}
\vspace{0.1cm}
\begin{tabular}{cccccccc}
\hline
& \multicolumn{2}{c}{Mean absolute error} & & & \multicolumn{2}{c}{Mean absolute error} \\
Parameter & Approx. exch. & Pseudo-post. & & Parameter & Approx. exch. & Pseudo-post. \\
\hline
$\beta_{1}$         & 0.469 & \textbf{0.409} & & $\gamma^*_{1,3}$  & \textbf{1.403} & 1.448 \\
$\beta_{2}$         & \textbf{0.384} & 0.505 & & $\gamma^*_{2,1}$  & \textbf{1.209} & 1.560 \\
$\beta^*_{1}$       & \textbf{0.308} & 0.326 & & $\gamma^*_{2,3}$  & 1.545 & \textbf{1.382} \\
$\beta^*_{2}$       & \textbf{0.335} & 0.435 & & $\gamma^*_{3,1}$  & \textbf{1.492} & 1.548 \\
$\gamma_{1,2}$      & \textbf{1.352} & 1.564 & & $\gamma^*_{3,2}$  & 1.883 & \textbf{1.428} \\
$\gamma_{1,3}$      & \textbf{1.707} & 1.784 & & $\delta_{1,2}$    & \textbf{0.382} & 0.596 \\
$\gamma_{2,1}$      & \textbf{1.413} & 1.649 & & $\delta_{1,3}$    & \textbf{0.343} & 0.421 \\
$\gamma_{2,3}$      & \textbf{1.530} & 1.566 & & $\delta_{2,1}$    & \textbf{0.387} & 0.649 \\
$\gamma_{3,1}$      & \textbf{1.467} & 1.594 & & $\delta_{2,3}$    & \textbf{0.398} & 0.503 \\
$\gamma_{3,2}$      & \textbf{1.390} & 1.515 & & $\delta_{3,1}$    & \textbf{0.305} & 0.540 \\
$\gamma^*_{1,2}$    & \textbf{1.368} & 1.415 & & $\delta_{3,2}$    & \textbf{0.444} & 0.604 \\
\hline
\end{tabular}
\label{tab7}
\end{center}
\end{table}

The approximate exchange consistently outperforms the pseudo-posterior method for the majority of the parameters, while the pseudo-posterior algorithm achieves a lower error in only a few cases, such as for $\beta_1$, $\gamma^*_{2,3}$, and $\gamma^*_{3,2}$. It is easy to note that the MAEs for the spatial parameters are relatively higher in Scenario D compared to Scenarios B and C. This can be imputed to the increased number of components, while the number of sites and time points remains similar to Scenario B and lower than in Scenario C. As a result, the number of observations per state is reduced, leading to greater variability and, consequently, higher MAEs.

Overall, the approximate exchange algorithm outperforms the pseudo-posterior approach across all scenarios considered. In particular, this behavior persists despite variations in spatial structure, number of sites, time points, and latent states, consistently resulting in lower MAEs.

\subsection{Synthetic data analysis}

We conclude the simulation study by examining two representative synthetic datasets, generated from Scenarios A and C, to illustrate the behavior of the two methods in greater detail. 

\subsubsection{Results for synthetic dataset 1}
\label{sec:dat1}

\begin{figure}[h]
\centering
\includegraphics[width=0.8\textwidth]{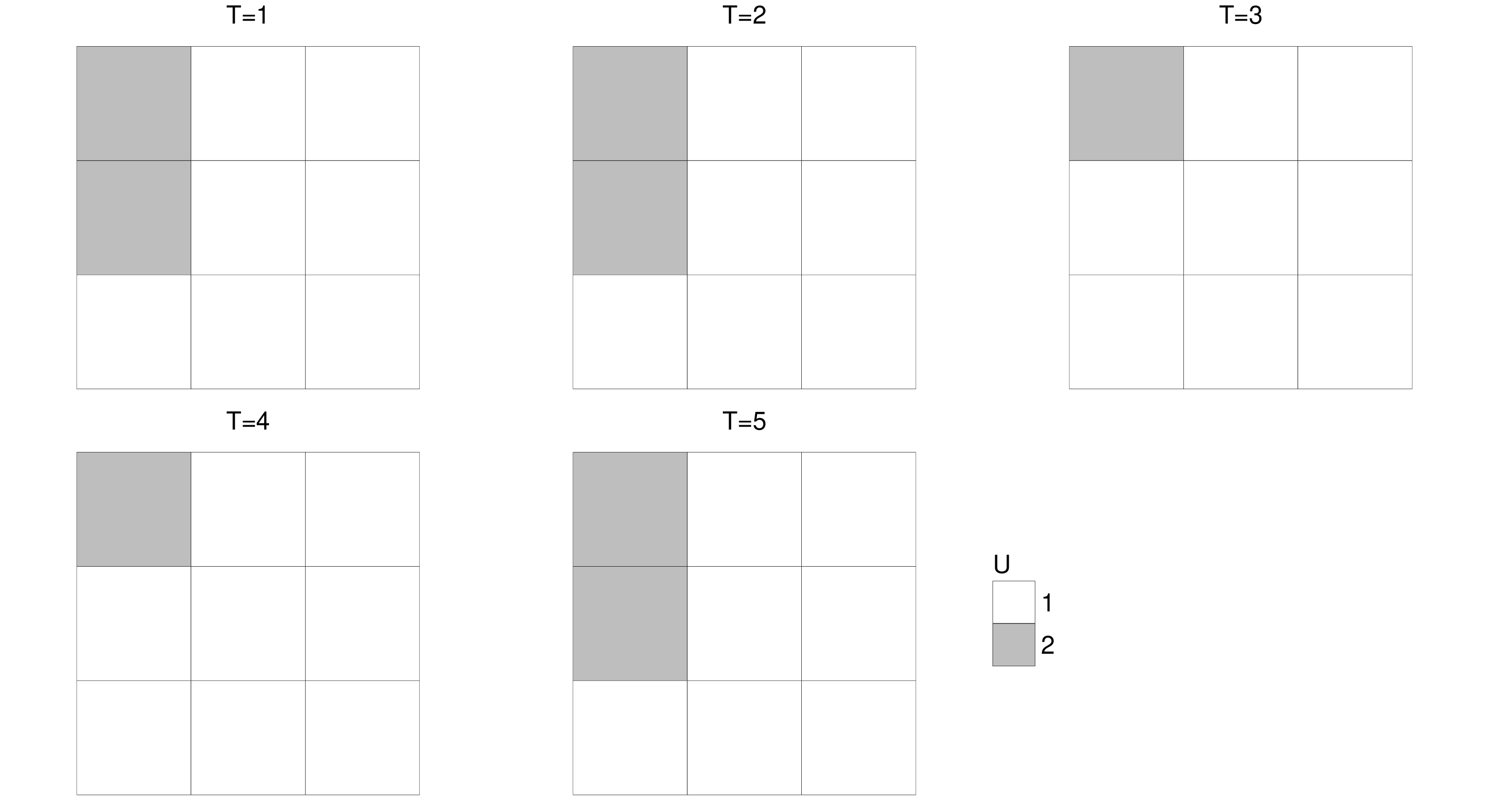} 
\caption{Synthetic data 1 generated following the setting defined in Scenario A.} 
\label{fig1}
\end{figure}

This dataset is generated following the setting provided in Scenario A. In particular, we fix $N=9$ and $T=5$, while for the number of states $K=2$ is used. A graphical representation of the latent process is provided in Figure \ref{fig1}, where a square grid $N=3\times3$ is defined, allowing to vary the clustering allocation over time. The generated dataset closely reflects patterns typically observed in real-world data, exhibiting a clear spatial dependence, where nearby locations tend to be more similar than distant ones, and a temporal persistence, where states are likely to remain stable over time. 

We estimate the spatio-temporal model defined in Section \ref{sec:model}, following the pseudo-posterior approach and the approximate exchange algorithm described in Section \ref{sec:inf}, according to the prior hyperparameters defined in Section \ref{sec:simdes}. The algorithms are run for 10,000 iterations and the first 5,000 are considered as initial burn-in, without considering any thinning. The convergence of each parameter is monitored using the Geweke test \citep{geweke1992evaluating} at a confidence level of 95\%, in the R package \textbf{coda}. The posterior expectation for each parameter, as well as the Monte Carlo standard error \citep{flegal2010batch} and the Geweke test, are reported in Table \ref{tab1}.
\begin{table}[h]
\small
\begin{center}\caption{Comparison between approximate
exchange and pseudo-posterior MCMC algorithms for synthetic data 1. In Geweke test, “Yes" means that the null hypothesis, that is mean estimates have converged, is not rejected.} 
{\begin{tabular}{clrrrrr}
\hline
& & \multicolumn{2}{c}{Posterior expectation (Monte Carlo s.e.)} & & \multicolumn{2}{c}{Geweke test}\\
 State & Parameter & Approx. Exchange & Pseudo-post. & True & Approx. Exchange & Pseudo-post. \\ 
\hline
 & & & & & & \\
 & & & & & & \\
\hline 
1 & $\mu_{1}$ & -3.117 (0.002) & -3.114 (0.002) & -3.0 & \text{Yes} & \text{Yes} \\ 
& $\mu_{2}$& -2.777 (0.002) & -2.777 (0.003) & -3.0 & \text{Yes} & \text{Yes} \\ 
 & $\sigma_{1,1}$& 1.000 (0.003) & 1.000 (0.003) & 1.0 & \text{Yes} & \text{Yes} \\ 
 & $\sigma_{1,2} $ & -0.334 (0.002) & -0.334 (0.003) & 0.0 & \text{Yes} & \text{Yes} \\ 
 & $\sigma_{2,2} $ & 1.217 (0.004) & 1.217 (0.004) & 1.0 & \text{Yes} & \text{Yes} \\ 
\hline
 2 & $\mu_{1} $& 2.858 (0.007) & 2.864 (0.007) & 3.0 & \text{Yes} & \text{Yes}\\ 
 & $\mu_{2} $ & 2.911 (0.006) & 2.900 (0.006) & 3.0 & \text{Yes} & \text{Yes}\\ 
 & $\sigma_{1,1}$ &1.881 (0.014) & 1.852 (0.013) & 1.0 & \text{Yes} & \text{Yes} \\ 
 & $\sigma_{1,2}$ & -0.556 (0.009) & -0.537 (0.008) & 0.0 & \text{Yes} & \text{Yes} \\ 
 & $\sigma_{2,2}$ & 1.603 (0.011) & 1.582 (0.011) & 1.0 & \text{Yes} & \text{Yes} \\ 
\hline 
 & $\beta_1$ & 0.836 (0.045) & 0.337 (0.054) & 2.0 &\text{Yes} & \text{No} \\ 
 & $\beta^*_1$ & 2.061 (0.085) & 0.418 (0.031) & 2.0 & \text{Yes} & \text{Yes} \\ 
 & $\gamma_{1,2}$ & -1.320 (0.045) & -0.278 (0.044) & -1.0 & \text{Yes} & \text{Yes} \\ 
 & $\gamma_{2,1}$ & 0.428 (0.046) & -0.132 (0.034) & 1.0 & \text{Yes} & \text{No} \\ 
 & $\gamma^*_{1,2}$ & -2.637 (0.095) & -0.250 (0.045) & -1.0 & \text{Yes} & \text{Yes} \\ 
 & $\gamma^*_{2,1}$ & 0.693 (0.059) & -0.163 (0.020) & 1.0 & \text{Yes} & \text{Yes} \\ 
 & $\delta_{1,2}$ & -1.357 (0.045) & -1.424 (0.029) & -1.0 & \text{Yes} & \text{Yes} \\ 
 & $\delta_{2,1}$ & -1.652 (0.035) & -1.671 (0.030) & -1.0 & \text{Yes} & \text{Yes} \\ 
\hline 
\end{tabular}}\label{tab1}
\end{center}
\end{table}

\begin{figure}[h!]
\centering
\includegraphics[width=1\textwidth]{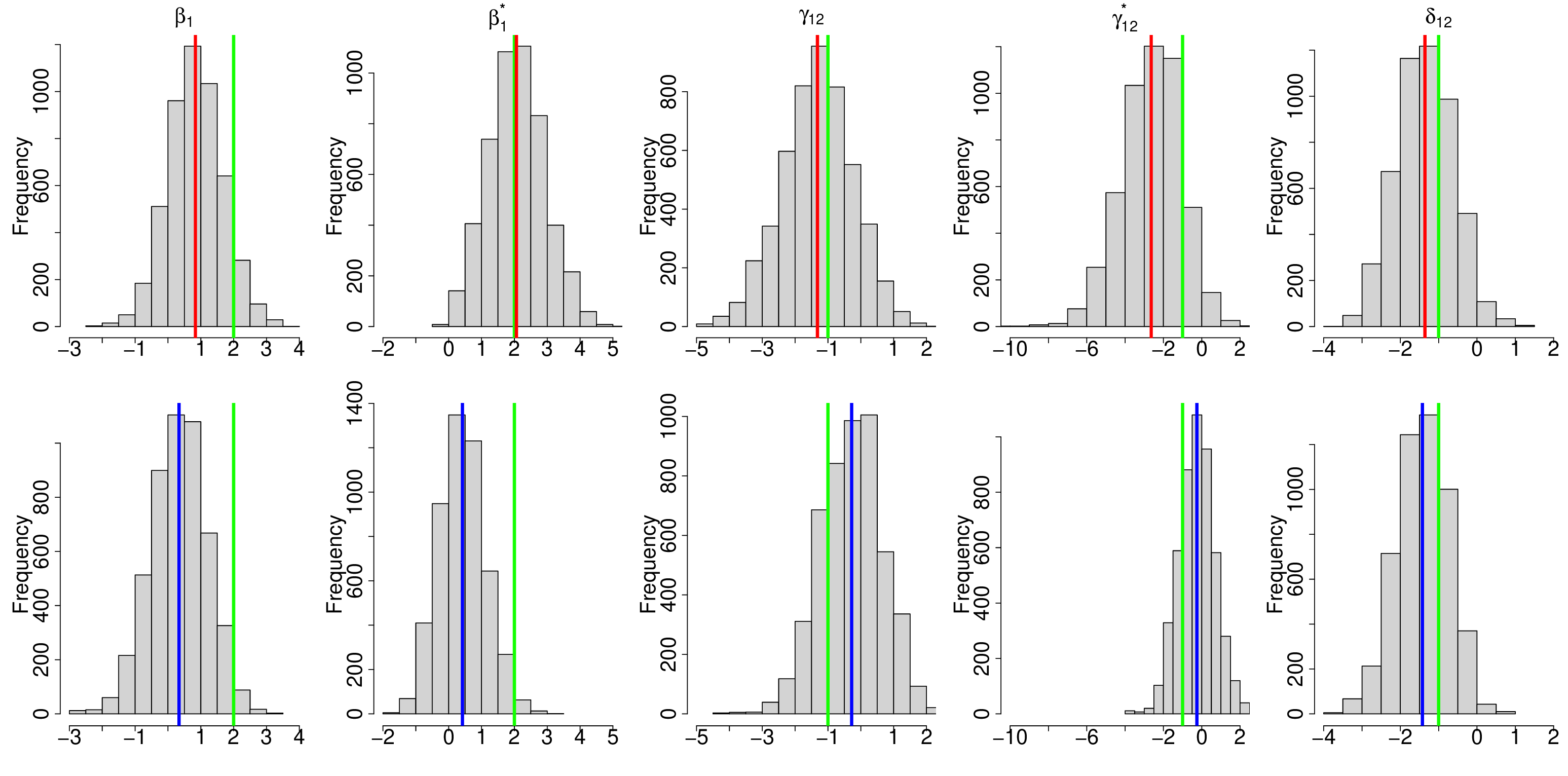} 
\caption{Histogram of the posterior samples obtained from the approximate exchange (top) and the pseudo-posterior algorithm (bottom) for the synthetic dataset 1. In green the true value of the generator, in red the posterior expectation of the exchange algorithm, and in blue the posterior expectation of the pseudo-posterior.} 
\label{fig2}
\end{figure}

A graphical representation for some of the parameters of the latent process are shown in Figure \ref{fig2}. 

Comparing the results obtained in Table \ref{tab1}, we note that for the parameters of the observable variables the results of both the algorithms are similar. The estimated parameters of the variance-covariance matrices exhibit a similar level of discrepancy from the true generating values across both methods. This deviation is likely attributable to the limited number of observations, which may reduce the precision of the estimates and may lead to greater variability in the inferred covariance structure. In particular, this deviation is observable for the state equal to 2, where there are only 8 latent variables in $K=2$ over the 45 latent variables involved. 

We observe that the approximate exchange algorithm generally provides more accurate estimates than the pseudo-posterior approach across the majority of parameters. However, when comparing the Monte Carlo standard errors, the pseudo-posterior method exhibits slightly better results if compared to the approximate exchange algorithm.

Furthermore, as illustrated by the histograms in Figure \ref{fig2}, the marginal posterior distributions obtained via the approximate exchange algorithm show higher variance compared to those produced by the pseudo-posterior method. This behavior is consistent with findings in the existing literature and is a known characteristic of the approximate exchange framework. A potential strategy to address this increased variability is the adoption of the noisy exchange algorithm, which introduces controlled noise to improve efficiency reducing the posterior variance \citep{alquier2016noisy}.

In addition, we evaluate the misclassification rate of the latent variables estimated and the real latent variables generated. The estimations are obtained using the MAP method. For both  algorithms, the misclassification rate is equal to 0, meaning that they easily identify the latent variables. This is something that we expect since the means of the 2 states are well separated and there is not a strong overlapping over the two components. 

Overall, the approximate exchange algorithm, when coupled with the proposed initialization strategy for the auxiliary component, appears to be a better alternative to the pseudo-posterior algorithm. This combined approach consistently yields more accurate and reliable results, particularly in estimating posterior expectations. 

\subsubsection{Results for synthetic dataset 2}

This dataset is generated following the setting provided in Scenario C, where $K=2$, $N=40$ and $T=10$. Also in this case, the response variable has a multivariate Gaussian distribution and the generated dataset reflects patterns typically observed in real-world data. The neighbourhood system is generated using the Erdős-Renyi model \citep{erdds1959random}, choosing uniformly at random from the collection of all graphs which have 
40 nodes and 20 edges. The dataset obtained is slightly imbalanced, since we observe 288 latent variables equal to 1 (72\%) and 112 latent variables equal to 2 (28\%), over 400 total latent variables. 

\begin{figure}[h!]
\centering
\includegraphics[width=1\textwidth]{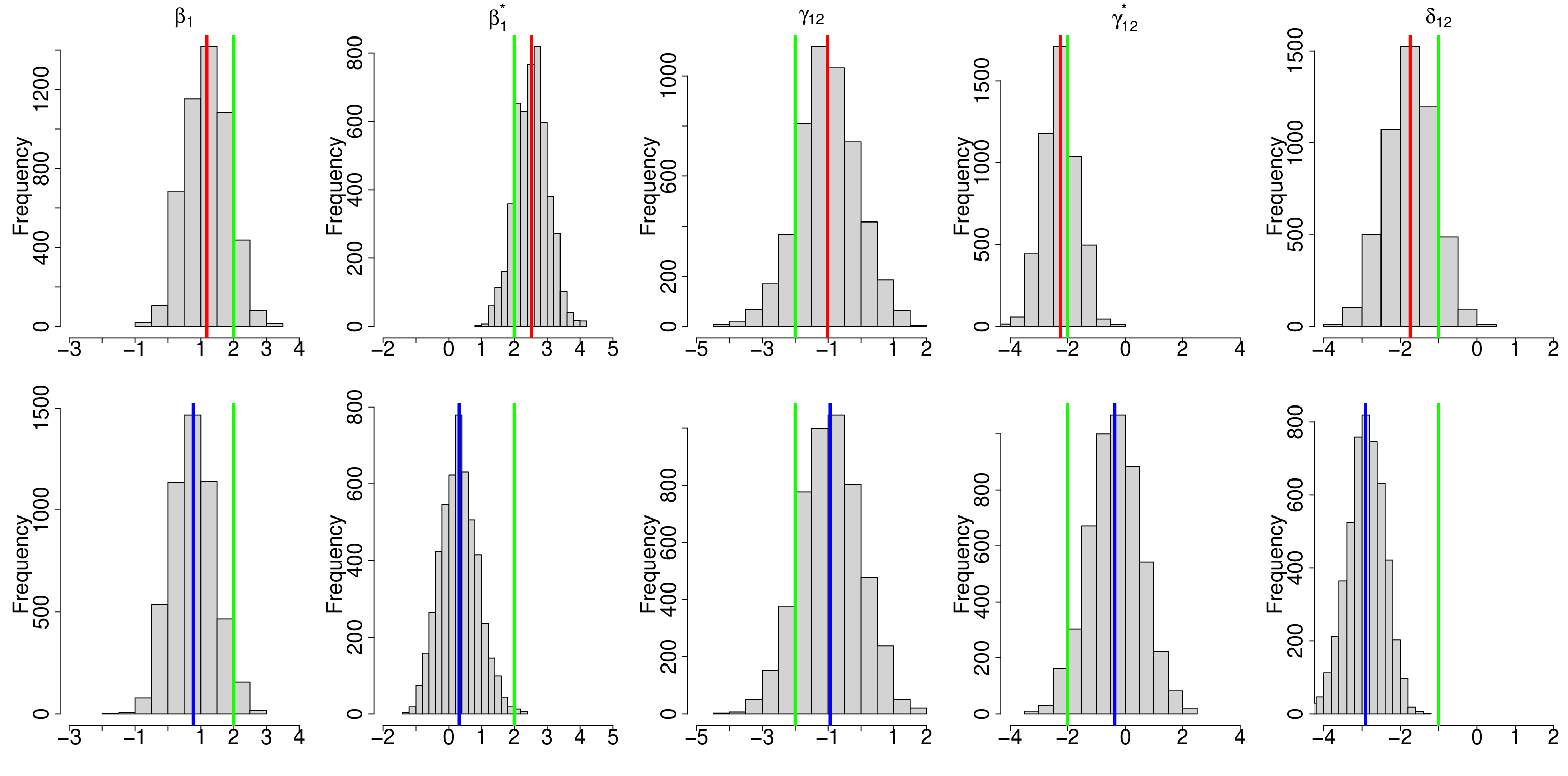} 
\caption{Histogram of the posterior samples obtained from the approximate exchange (top) and the pseudo-posterior algorithm (bottom) for the synthetic dataset 2. In green the true value of the generator, in red the posterior expectation of the exchange algorithm, and in blue the posterior expectation of the pseudo-posterior.} 
\label{fig3}
\end{figure}

We estimate the same model defined in the previous section, running for 10,000 iterations the algorithms and considering the first 5,000 as initial burn-in, without any thinning. The convergence of each parameter is monitored using the Geweke test comparing then the posterior expectation for each parameter, as well as the Monte Carlo standard error. We report the results in Table \ref{tab2} and a graphical representation of same samples of the latent parameters is shown in Figure \ref{fig3}.
\begin{table}[h]
\small
\begin{center}\caption{Comparison between approximate
exchange and pseudo-posterior MCMC algorithms for synthetic data 2. In Geweke test, “Yes" means that the null hypothesis, that is mean estimates have converged, is not rejected.} 
{\begin{tabular}{clrrrrr}
\hline
& & \multicolumn{2}{c}{Posterior expectation (Monte Carlo s.e.)} & & \multicolumn{2}{c}{Geweke test}\\
 State & Parameter & Approx. Exchange & Pseudo-post. & True & Approx. Exchange & Pseudo-post. \\ 
\hline
 & & & & & & \\
 & & & & & & \\
\hline 
1 & $\mu_{1}$ & -3.049 (0.001) & -3.050 (0.001) & -3.0 & \text{Yes} & \text{Yes} \\ 
& $\mu_{2}$& -2.920 (0.001) & -2.921 (0.001) & -3.0 & \text{Yes} & \text{Yes} \\ 
 & $\sigma_{1,1}$& 0.903 (0.001) & 0.902 (0.002) & 1.0 & \text{Yes} & \text{Yes} \\ 
 & $\sigma_{1,2} $ & 0.023 (0.001) & 0.025 (0.001) & 0.0 & \text{Yes} & \text{Yes} \\ 
 & $\sigma_{2,2} $ & 1.090 (0.001) & 1.090 (0.001) & 1.0 & \text{Yes} & \text{Yes} \\ 
\hline
 2 & $\mu_{1} $& 2.948 (0.001) & 2.944 (0.001) & 3.0 & \text{Yes} & \text{Yes}\\ 
 & $\mu_{2} $ & 2.929 (0.001) & 2.930 (0.001) & 3.0 & \text{Yes} & \text{Yes}\\ 
 & $\sigma_{1,1}$ &1.029 (0.002) & 1.035 (0.001) & 1.0 & \text{Yes} & \text{Yes} \\ 
 & $\sigma_{1,2}$ & -0.109 (0.001) & -0.112 (0.001) & 0.0 & \text{Yes} & \text{Yes} \\ 
 & $\sigma_{2,2}$ & 0.825 (0.011) & 0.825 (0.001) & 1.0 & \text{Yes} & \text{Yes} \\ 
\hline 
 & $\beta_1$ & 1.186 (0.045) & 0.764 (0.037) & 2.0 &\text{Yes} & \text{Yes} \\ 
 & $\beta^*_1$ & 2.521 (0.061) & 0.315 (0.021) & 2.0 & \text{Yes} & \text{Yes} \\ 
 & $\gamma_{1,2}$ & -1.013 (0.054) & -0.937 (0.041) & -2.0 & \text{Yes} & \text{Yes} \\ 
 & $\gamma_{2,1}$ & 1.597 (0.041) & 0.715 (0.029) & 2.0 & \text{Yes} & \text{Yes} \\ 
 & $\gamma^*_{1,2}$ & -2.253 (0.056) & -0.355 (0.045) & -2.0 & \text{Yes} & \text{Yes} \\ 
 & $\gamma^*_{2,1}$ & 3.193 (0.061) & 0.124 (0.024) & 2.0 & \text{Yes} & \text{Yes} \\ 
 & $\delta_{1,2}$ & -1.737 (0.047) & -2.904 (0.016) & -1.0 & \text{Yes} & \text{Yes} \\ 
 & $\delta_{2,1}$ & -1.360 (0.046) & -1.874 (0.033) & -1.0 & \text{Yes} & \text{Yes} \\ 
\hline 
\end{tabular}}\label{tab2}
\end{center}
\end{table}

We evaluate the misclassification rate using the MAP method for the estimation of the latent variables. As in Section \ref{sec:dat1}, the misclassification rate is equal to 0 since the means of the two states are well separated and there is not strong overlapping over the two components. Comparing the results obtained in Table \ref{tab2} and Figure \ref{fig3}, we note that for the parameters of the response variable the results for both the algorithms are similar. The estimated parameters of the variance-covariance matrices exhibit a lower level of discrepancy from the true generating values across both methods if compared to the previous section. This reduction is attributable to the increasing number of observations. Looking at the latent variable parameters, it is possible to see that the approximate exchange algorithm generally delivers more accurate estimates than the pseudo-posterior approach for most parameters. However, when evaluating the Monte Carlo standard errors, the pseudo-posterior method shows slightly better performance compared to the approximate exchange algorithm, as in the previous simulation study. 

In conclusion, as obtained in Section \ref{sec:dat1}, the approximate exchange algorithm seems to be a better alternative to the pseudo-posterior algorithm. The approximate exchange approach produces more accurate outcomes, especially in the estimation of posterior expectations, highlighting its effectiveness in enhancing the overall quality of inference.

\section{Application}\label{sec:apl}

In this section, we present an application of the proposed model to the analysis of meteorological trends in Italy, focusing specifically on regional-level precipitation data. These data are available at the official page of the Italian National Institute of Statistics (ISTAT)\footnote{\url{https://www.istat.it/comunicato-stampa/andamento-meteo-climatico-in-italia-anni-2000-2009/}}.  

In this dataset, ISTAT presents key findings on meteorological trends in Italy. The data analysis is based on observations from approximately 150 meteorological stations, carried out in collaboration with the Council for Agricultural Research and Analysis of the Agricultural Economy - Research Unit for Climatology and Meteorology Applied to Agriculture (CRA-CMA).

\begin{figure}[h!]
\centering
\includegraphics[width=1\textwidth]{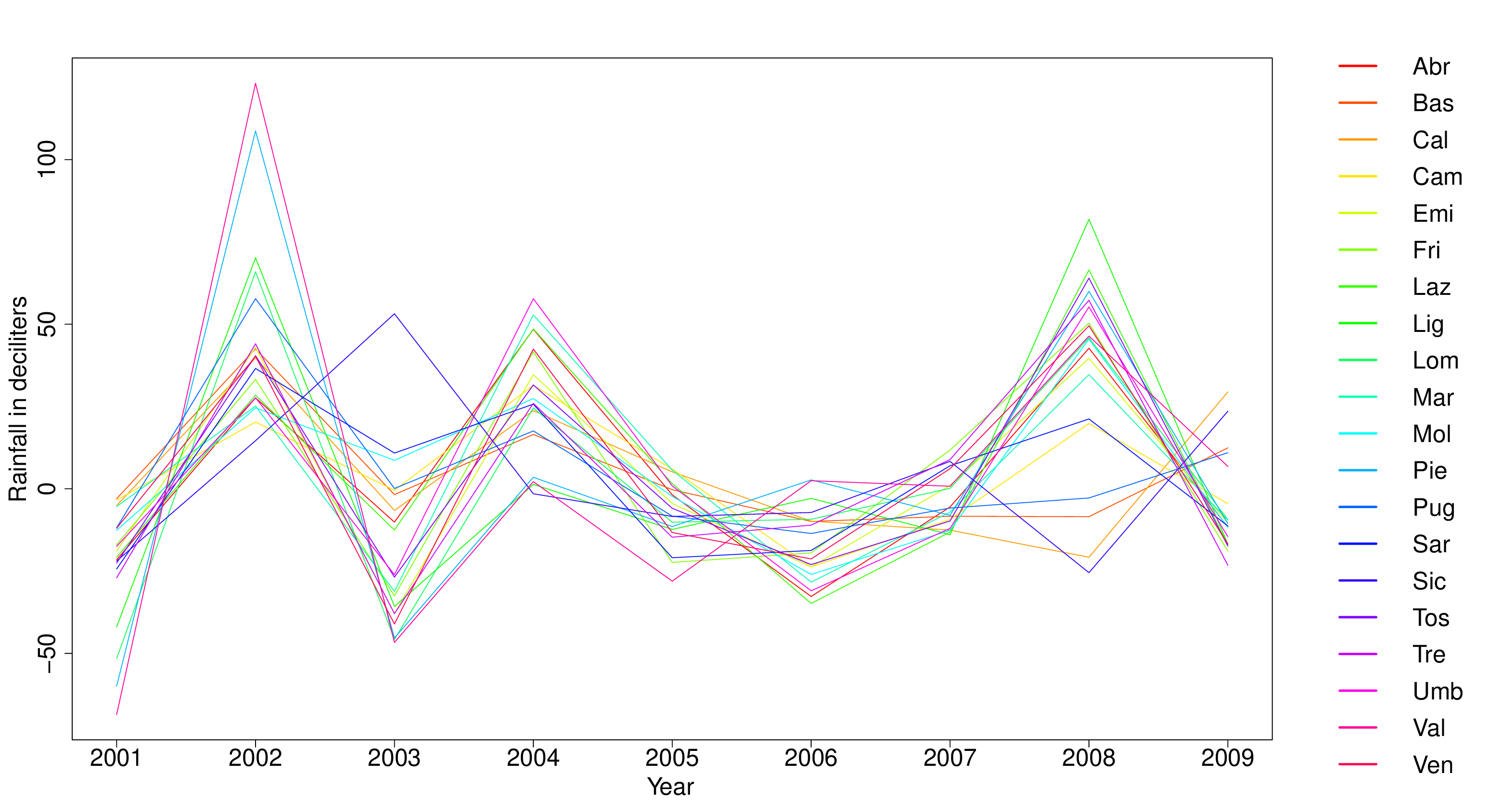} 
\caption{Multi-series visualization of regional rainfall data expressed in liters for the period 2000-2009.} 
\label{fig5}
\end{figure}

The dataset covers the decade from 2000 to 2009 and includes annual data on temperature and precipitation, with territorial detail at the national, macro-regional, regional, and provincial levels. In particular, the dataset used concerns the 20 Italian regions and the weighted average of the yearly rainfalls expressed in liters, computed by ISTAT using the surface area of each individual region as weights. A graphical representation of the dataset is reported in Figure \ref{fig5}.

\begin{figure}[h!]
\centering
\includegraphics[width=1\textwidth]{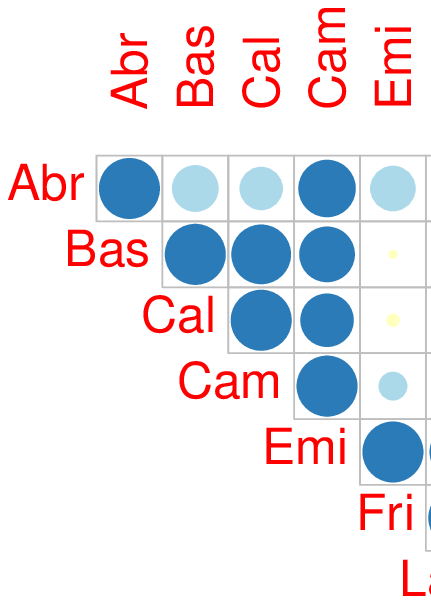} 
\caption{Correlation among the different regions for the rainfall evaluated in liters, for the period 2000-2009.} 
\label{fig4}
\end{figure}

A summary of the dataset is reported in Table \ref{tab8}. In addition, it is possible to visualize possible correlations among the regions, and these values are reported in Figure \ref{fig4}.
\begin{table}[h!]
\centering
\caption{Average rainfall values for each Italian region expressed in liters for the period 2000-2009.}
\vspace{0.1cm}
\begin{tabular}{l c l c}
\hline
Region & Mean & Region & Mean \\
\hline
Abruzzo (Abr)     & 0.810 & Liguria (Lig)     & 0.807 \\
Basilicata (Bas)  & 0.702 & Lombardia (Lom)   & 0.829 \\
Calabria (Cal)    & 0.767 & Marche (Mar)      & 0.755 \\
Campania (Cam)    & 0.779 & Molise (Mol)      & 0.752 \\
Emilia-Romagna (Emi) & 0.766 & Piemonte (Pie)    & 0.845 \\
Friuli-Venezia Giulia (Fri) & 1.073 & Puglia (Pug)       & 0.626 \\
Lazio (Laz)       & 0.802 & Sardegna (Sar)    & 0.496 \\
Sicilia (Sic)     & 0.620 & Toscana (Tos)     & 0.756 \\
Trentino-Alto Adige (Tre) & 0.814 & Umbria (Umb)      & 0.800 \\
Valle d'Aosta (Val) & 0.846 & Veneto (Ven)     & 0.859 \\
\hline
\end{tabular}
\label{tab8}
\end{table}

From the preliminary analysis of the data, we observe distinct regional patterns in the rainfall distribution across Italy. The northern regions, such as Friuli-Venezia Giulia, Trentino-Alto Adige, Valle d'Aosta, Liguria, Lombardia, Piemonte and Veneto, generally have higher levels of rainfall in comparison to those in the central and southern parts of the country. Furthermore, the correlation between the rainfall series from different regions is often strongly positive. However, few exceptions to this pattern can be observed, for example when we compare regions that are geographically distant from each other. 

We initially analyze the original dataset using the model defined in Section \ref{sec:model}, selecting the number of components based on the Deviance Information Criterion (DIC) \citep{spiegelhalter2002bayesian}. Due to the characteristic of the data, the selected model is a spatio-temporal model with two latent states, identifying periods of high and low rainfall. In addition, for several years the model identifies only a single cluster, yielding results that align closely with those from a preliminary exploratory analysis. In addition, given that the rainfall values are non-negative and our response variable follows a Gaussian distribution, instead of using the original dataset, we analyze the relative variation in rainfall, defined as$$y_{i,t} = \frac{r_{i,t} - r_{i,t-1}}{r_{i,t-1}} \times 100,$$
where $r_{i,t}$ is the rainfall for a region $i$ at time $t$. Notice that $y_{i,t}$ can take both positive and negative values.
In detail, we use as a response variable an univariate Gaussian distribution and as prior distributions
$$
\mu_u \sim \mathcal{N}(0, 1000) \quad \text{and} \quad \sigma_u \sim \mathcal{IG}(2, 1), \quad u=1,\ldots,K,
$$
where $\mathcal{IG}(\cdot, \cdot)$ denotes an Inverse-gamma. The full conditional distributions, which includes also the Inverse-gamma prior distribution, are reported in Appendix C. For this application 50,000 iterations are considered, with 10,000 iterations of burn-in and a thinning of 10 iterations. Diagnostic analysis are performed, using as for the simulation study the Geweke test \citep{geweke1992evaluating}.

Model selection, specifically determining the number of latent states, is carried out using the DIC, considering the formula which selects the lowest value of DIC as optimal. We start with a model that includes only one latent state, we calculate the DIC, and then progressively increase the number of states, evaluating the DIC at each step. This process continues until we observe that the DIC shifts from a lower value to a higher one, and at that point, we stop and we select the model with lower DIC value. The results obtained are reported in Table \ref{tab9}.
\begin{table}[h!]
\centering
\caption{DIC values for models with different numbers of latent states.}
\vspace{0.1cm}
\begin{tabular}{c c}
\hline
Number of states $K$ & DIC \\
\hline
1 & $1750.363$ \\
2 & $1637.357$ \\
\textbf{3} & $\textbf{1608.102}$ \\
4 & $1608.802$ \\
\hline
\end{tabular}
\label{tab9}
\end{table}

\begin{figure}[h!]
\centering
\includegraphics[width=1\textwidth]{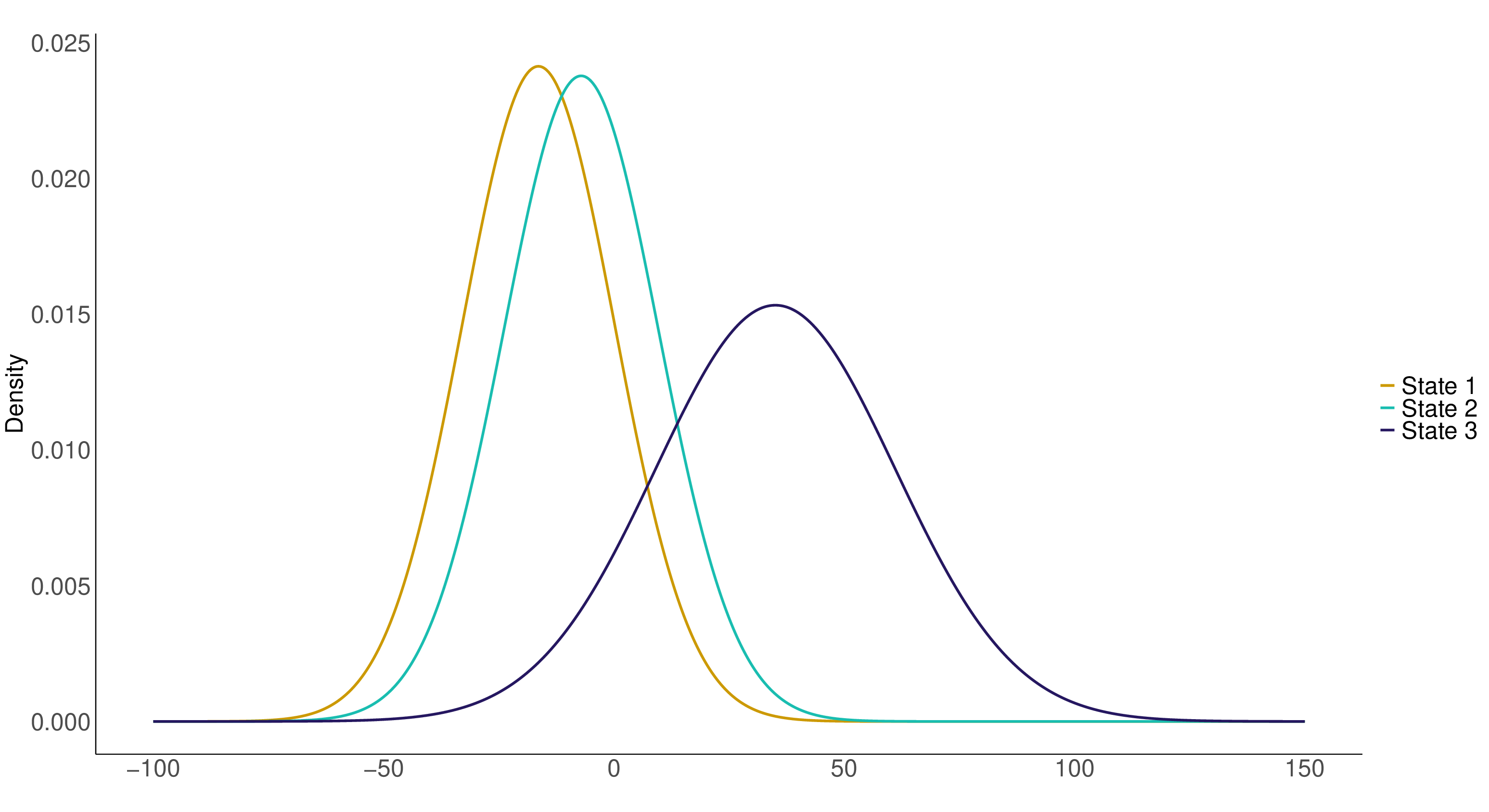} 
\caption{Gaussian densities associated to the three different states obtained from the model.} 
\label{fig6}
\end{figure}

As it is clear from Table \ref{tab9}, the final model selected is that $K=3$, which corresponds the smallest DIC among the other models. The means and variances estimated are 
$$\hat{\mu}_1 = -16.382, \quad \hat{\mu}_2 =  -7.106 \quad \text{and} \quad \hat{\mu}_3 = 35.069,$$
while
$$\hat{\sigma}_1 = 16.531, \quad \hat{\sigma}_2 = 16.776 \quad \text{and} \quad \hat{\sigma}_3 = 26.021.$$
A graphical representation of the three Gaussian densities obtained, respectively for state 1, 2 and 3, is reported in Figure \ref{fig6}.

It is possible to note that the densities of state 1 and state 2 slightly tend to overlap with each other, identifying a less separation between these two states. 
For the latent variables, we have the following results
$$\hat{\boldsymbol{\beta}} = (0.923,-0.010,0)' \quad \text{and} \quad \hat{\boldsymbol{\beta}^*} = (-0.035,0.235,0)',$$
while for the spatial parameters we have 
$$
\hat{\boldsymbol{\gamma}} = \begin{pmatrix}
0 & 0.359 & -0.643 \\
0.451 & 0  & -0.329  \\
-0.546 & -0.187 & 0 & \\
\end{pmatrix},
\quad
\hat{\boldsymbol{\gamma}^*} = \begin{pmatrix}
0 & 1.407 & -5.215 \\
1.694 & 0  & -5.087 \\
-4.202 & -4.359 & 0\\
\end{pmatrix},
$$
and for the temporal parameters
$$
\hat{\boldsymbol{\delta}} = \begin{pmatrix}
0 & -0.281 & 1.523 \\
-0.250 & 0  & 0.181  \\
0.547 & 0.335 & 0 & \\
\end{pmatrix}.
$$
The estimated parameters for $\boldsymbol{\beta}$ suggest a high probability of observing the first state. In contrast, $\boldsymbol{\beta}^*$ indicates a slightly preference of the second state, with the first state showing a value close to zero. The spatial parameters in $\boldsymbol{\gamma}$ and $\boldsymbol{\gamma}^*$ produce similar outcomes, identifying clusters that align with neighboring regions. Specifically, when $u = 1$ and $u = 2$, the neighboring sites of $i$ exhibit similar states. This is due to the negative values of $\gamma_{1,3}$, $\gamma_{2,3}$, $\gamma_{3,1}$, and $\gamma_{3,2}$, as well as in $\boldsymbol{\gamma}^*$. Positive values are obtained for $\gamma_{1,2}$ and $\gamma_{2,1}$, as well as $\gamma^*_{1,2}$ and $\gamma^*_{2,1}$, indicating a propension of clusters 1 and 2 to appear as neighbours rather often. Regarding the temporal dynamics, the parameters show a slightly high probability of moving out of cluster 3, but also a noticeably high probability of moving into cluster 3, from either cluster 1 or 2. We interpret this as cluster 3 being a rather recurrent state where nodes tend not to stay for particularly long.
The three different latent states can identify 3 different relative variation levels of rainfall, and based on the $\mu_u$ obtained, we can identify a low, medium-low and high level of relative variances. 

\begin{figure}[h!]
\centering
\includegraphics[width=1\textwidth]{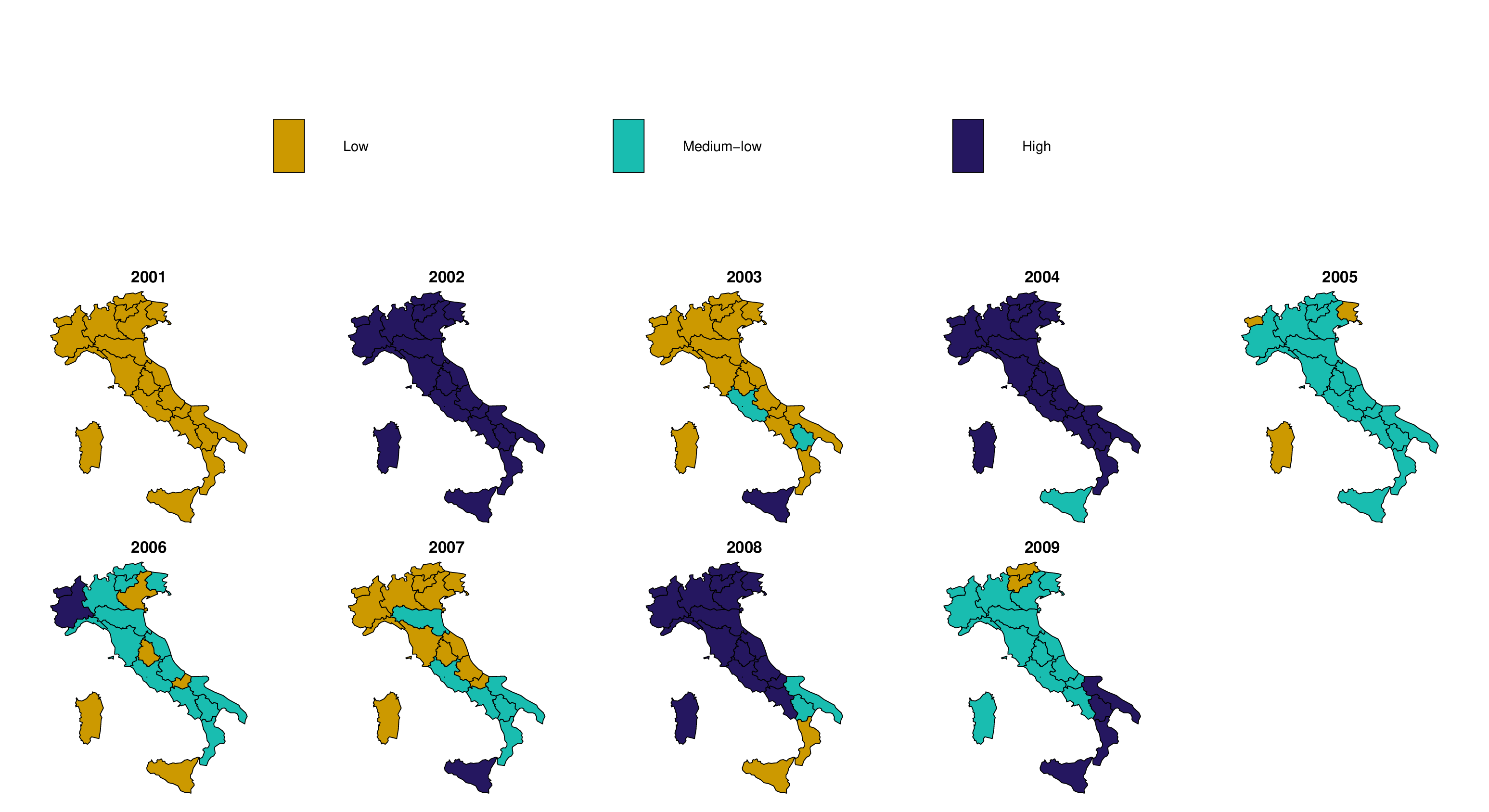} 
\caption{Spatio-temporal cluster associated to the 3 different states for each region in Italy across the years from 2001 to 2009.} 
\label{fig7}
\end{figure}

Since we use a data augmentation approach, the latent variables can be estimated using a MAP approach, which leads to a clear graphical representation shown in Figure \ref{fig7}.

We can highlight some geographical and temporal patterns in the relative variation rainfall distribution. First, a consistent level of relative variations across all regions is evident in 2001 and 2002. However, this uniformity does not emerge in the following years. Second, a clear spatial dependence persists over time, with the identification of clusters composed of neighboring regions exhibiting similar levels of relative variation rainfalls. This pattern is identified by the model, as shown in the estimated parameters for spatial dependencies. A typical tendency to observe extreme variation over years is notable, comparing for example the first four years and the last three years. This pattern is also captured by the model, as shown in the estimated parameters for temporal dependencies. Specifically, in state $u=2$, we observe a persistence of medium-low variation levels over time. In contrast, the other states tend to transition between $u=1$ and $u=3$, as well as $u=3$ and $u=1$, indicating greater variability. 

\begin{figure}[h!]
\centering
\includegraphics[width=1\textwidth]{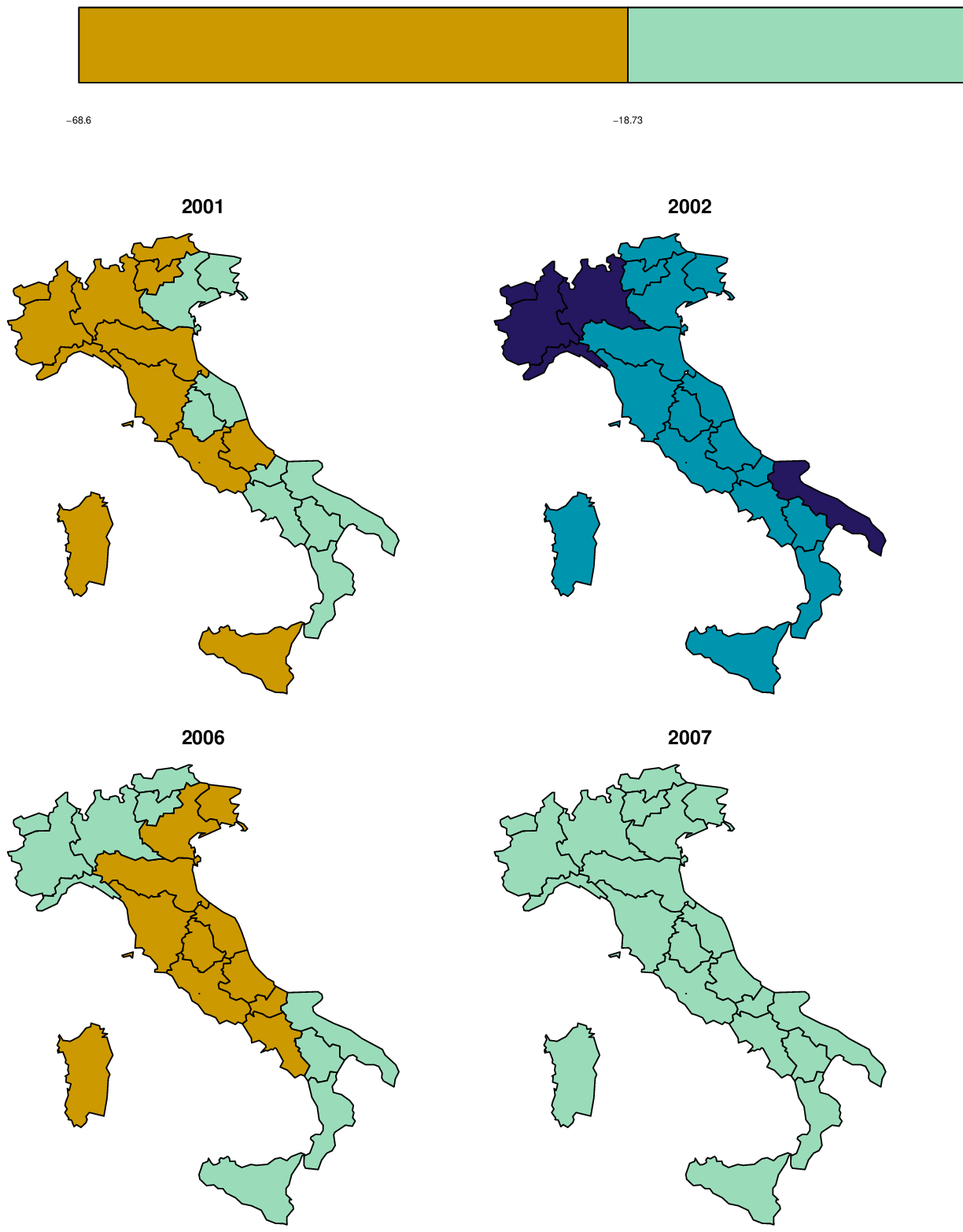} 
\caption{Rainfall variations divided by quartile for each region in Italy across the years from 2001 to 2009.} 
\label{fig:prel}
\end{figure}

A preliminary analysis also highlights similar patterns, based on a map of Italian regions grouped into four quartiles according to changes in rainfall. This is visually illustrated in Figure \ref{fig:prel}, Appendix D. As expected, comparable trends emerge, though some regional differences are evident, for instance, Piemonte in 2006, or Puglia and Basilicata in 2009. Differently from the preliminary analysis, the proposed model offers a more structured interpretation of these phenomena through its set of parameter. Moreover, the number of clusters is statistically validated using a specific criterion.

In conclusion, Figure \ref{fig7} and the estimated parameters of the model proposed identify specific patterns, reflecting increased variability in rainfall levels, in particular the shifts between wet and dry periods over time. These characteristics were already examined in \cite{tsonis1996widespread}, where an analysis over 5,328 stations around the globe up to the late 1980s was considered. More recently, further evidence of the amplification of precipitation variability has been presented in a study published in Science \citep{zhang2024anthropogenic}.

\section{Conclusions}

We propose a spatio-temporal hidden Markov model that is flexible and can be adapted to various types of response variables. Our new framework extends the models proposed in \cite{bartolucci2022hidden} and \cite{bartolucci2022spatio} to include individual initial time parameters and specific parameters for the prevalence of single and transition-states, for both spatial and temporal components.

Focusing on the Bayesian estimation of the proposed model, we have introduced an approximate exchange algorithm that improves upon the classical pseudo-posterior approach commonly used in the context of spatio-temporal hidden Markov models. The proposed algorithm directly targets the true posterior distribution in the Markov chain Monte Carlo algorithm, avoiding the use of pseudo-distributions. The algorithm requires to augment the posterior distribution including an auxiliary process which has to be equal in distribution to the intractable one. For the sampling of the auxiliary process a Gibbs sampler is used, while to address the computational cost typically associated with the exchange algorithm, we introduce an alternative initialization strategy for the auxiliary variable. This refinement reduces the number of iterations needed for the auxiliary variable, thereby improving overall the computational time.

We assess the performance of the proposed method through both simulated and real data applications. In particular, we compare the approximate exchange approach with the standard pseudo-posterior method across various scenarios, varying the spatial structures, the number of sites, time occasions, and the number of latent states. Across all scenarios, the proposed approximate exchange demonstrates consistently positive results, outperforming the pseudo-posterior algorithm.

Future work will focus on the scalability of the algorithm for high-dimensional datasets and exploring alternative Markov chain Monte Carlo techniques suited for this model class. 

\section*{Appendix A}
\label{apA}

In this appendix we prove that the model defined in Equation \eqref{eq:lat} satisfies the property in Equation \eqref{eq:dip}. Let $t>1$, we have that 

$$p(U_{i,t} = k \lvert \boldsymbol{U}_{-(i,t)} = \boldsymbol{u}_{-(i,t)}, \boldsymbol{\theta}) = \frac{p(U_{i,t} = k, \boldsymbol{U}_{-(i,t)} = \boldsymbol{u}_{-(i,t)} \lvert \boldsymbol{\theta})}{p(\boldsymbol{U}_{-(i,t)} = \boldsymbol{u}_{-(i,t)} \lvert \boldsymbol{\theta})}.$$
Based on Equation \eqref{eq:ratio}, we have 

\begin{equation*}
\begin{split}
\frac{p(U_{i,t} = k, \boldsymbol{U}_{-(i,t)} = \boldsymbol{u}_{-(i,t)} \lvert \boldsymbol{\theta})}{p(\boldsymbol{U}_{-(i,t)} = \boldsymbol{u}_{-(i,t)} \lvert \boldsymbol{\theta})} &= \frac{q_{\boldsymbol{\theta}}(U_{i,t} = k, \boldsymbol{U}_{-(i,t)} = \boldsymbol{u}_{-(i,t)})/\mathcal{Z}_{\boldsymbol{\theta}}}{\sum_{z=1}^K q_{\boldsymbol{\theta}}(U_{i,t} = z, \boldsymbol{U}_{-(i,t)} = \boldsymbol{u}_{-(i,t)})/\mathcal{Z}_{\boldsymbol{\theta}}}\\
&= \frac{q_{\boldsymbol{\theta}}(U_{i,t} = k, \boldsymbol{U}_{-(i,t)} = \boldsymbol{u}_{-(i,t)})}{\sum_{z=1}^K q_{\boldsymbol{\theta}}(U_{i,t} = z, \boldsymbol{U}_{-(i,t)} = \boldsymbol{u}_{-(i,t)})}.
\end{split}
\end{equation*}
Now, notice that we can simplify this ratio, due to the exponential form and considering only the $U_{i,t}$ term, obtaining
$$\frac{\exp{ \bigg\{ \beta_k^* + \gamma_{u_{i,t-1},k} + \sum_{\substack{j= i+1 \\ j \in \eta_i}}^{N} \sum_{v \neq k}   \mathds{1}(U_{i,t} = k, U_{j,t} = v) \gamma^*_{k,v} \bigg\} }}{\sum_{z=1}^K \exp{ \bigg\{ \beta_z^* + \gamma_{u_{i,t-1},z} + \sum_{\substack{j=i+1 \\ j \in \eta_i}}^{N} \sum_{v \neq z}   \mathds{1}(U_{i,t} = z, U_{j,t} = v) \gamma^*_{z,v} \bigg\} }},$$
which depends only on $U_{i,t-1}$ and $\boldsymbol{U}_{j \in \eta_i,t}$, as required.

\section*{Appendix B}
\label{apB}

In this appendix we show how to derive the full conditional in Equation \eqref{eq:mean} and \eqref{eq:variance}. Under the assumptions defined in Section 2.1, the full conditional for the mean ${\boldsymbol{\mu}_u}$, where $u=1,\ldots,K$, is obtained as follows:

\begin{equation}
\begin{aligned}
p({\boldsymbol{\mu}_u} \lvert \cdots)\propto & \prod_{i=1}^{N}\prod_{t=1}^T e^{-\frac{1}{2} (\boldsymbol{y}_{i,t}-\boldsymbol{\mu}_u)'\boldsymbol{\Sigma}_u^{-1}(\boldsymbol{y}_{i,t}-\boldsymbol{\mu}_u) \mathds{1}_{U_{i,t}}(u)}  e^{-\frac{1}{2} (\boldsymbol{\mu}_u - \boldsymbol{m})' \boldsymbol{V}^{-1} (\boldsymbol{\mu}_u - \boldsymbol{m})} \\
= &~ e^{-\frac{1}{2} \sum_i \sum_t (\boldsymbol{y}_{i,t}-\boldsymbol{\mu}_u)'\boldsymbol{\Sigma}_u^{-1}(\boldsymbol{y}_{i,t}-\boldsymbol{\mu}_u)\mathds{1}_{U_{i,t}}(u)}e^{-\frac{1}{2} (\boldsymbol{\mu}_u - \boldsymbol{m})' \boldsymbol{V}^{-1} (\boldsymbol{\mu}_u - \boldsymbol{m})} \\ 
 = & ~ e^{-\frac{1}{2} \left( \sum_i \sum_t \boldsymbol{y}_{i,t}'\boldsymbol{\Sigma}_u^{-1} \boldsymbol{y}_{i,t} \mathds{1}_{U_{i,t}}(u) - 2\sum_i \sum_t \boldsymbol{\mu}_u'\boldsymbol{\Sigma}_u^{-1} \boldsymbol{y}_{i,t} \mathds{1}_{U_{i,t}}(u)  + \sum_i \sum_t  \boldsymbol{\mu}_u'\boldsymbol{\Sigma}_u^{-1} \boldsymbol{\mu}_u \mathds{1}_{U_{i,t}}(u) \right) }\\ 
  & \times  e^{-\frac{1}{2} \left( \boldsymbol{\mu}_u' \boldsymbol{V}^{-1} \boldsymbol{\mu}_u -2 \boldsymbol{\mu}_u' \boldsymbol{V}^{-1} \boldsymbol{m} + \boldsymbol{m}' \boldsymbol{V}^{-1} \boldsymbol{m}\right) } \\
 \propto & ~ e^{ -\frac{1}{2} \left[ \boldsymbol{\mu}_u'(n_u\boldsymbol{\Sigma}_u^{-1} + \boldsymbol{V}^{-1})\boldsymbol{\mu}_u - 2 \boldsymbol{\mu}_u'(\boldsymbol{\Sigma}_u^{-1} n_u {\boldsymbol{\bar{y}}_u} + \boldsymbol{V}^{-1}\boldsymbol{m})\right] },\label{a1}
\end{aligned}
\end{equation}
where $\mathds{1}_{U_{i,t}}(u) = \mathds{1}(U_{i,t}=u)$,
$$n_u = \sum_{i=1}^{N} \sum_{t = 1}^T \mathds{1}_{U_{i,t}}(u), ~~~ \text{and}~~~ {\boldsymbol{\bar{y}}_u} = (1/n_u) \sum_{i=1}^N \sum_{t=1}^{T} \boldsymbol{y}_{i,t} \mathds{1}_{U_{i,t}}(u).$$ From Equation \eqref{a1} it is possible to recognize the Gaussian kernel. Under the same assumptions defined in Section 2.1, the full conditional for the variance-covariance matrix ${\boldsymbol{\Sigma}_u}$, with $u=1,\ldots,K$, is

\begin{equation}
\begin{aligned}
p({\boldsymbol{\Sigma}_u} \lvert \cdots)\propto & ~ \lvert \boldsymbol{\Sigma}_u \lvert ^{-n_u/2} 
e^{-\frac{1}{2} \sum_i \sum_t (\boldsymbol{y}_{i,t}-\boldsymbol{\mu}_u)'\boldsymbol{\Sigma}_u^{-1}(\boldsymbol{y}_{i,t}-\boldsymbol{\mu}_u) \mathds{1}_{U_{i,t}}(u)} \lvert \boldsymbol{\Sigma}_u \lvert ^{-(\nu  + d + 1)/2}  e^{-\frac{1}{2}{\rm tr}
 (\boldsymbol{S}  \boldsymbol{\Sigma}_u^{-1})} \\
= & ~ \lvert \boldsymbol{\Sigma}_u \lvert ^{-(\nu + n_u + d + 1)/2} e^{-\frac{1}{2} {\rm tr}
 \left\{\left[\boldsymbol{S} + \sum_i \sum_t (\boldsymbol{y}_{i,t} -\boldsymbol{\mu}_u)(\boldsymbol{y}_{i,t}-\boldsymbol{\mu}_u)'  \mathds{1}_{U_{i,t}}(u)\right]\boldsymbol{\Sigma}_u^{-1}\right\}},\label{a2} 
\end{aligned}
\end{equation}
where $\rm tr(\cdot)$ is the trace operator. From Equation \eqref{a2} it is possible to recognize the Inverse-Wishart kernel.

\section*{Appendix C}

Let $\mu_u \sim\mathcal{N}(m, v)$ and $\sigma_k^2 \sim\mathcal{IG}(a, b)$ for all $u=1,\ldots,K$. We have

\begin{equation*}
\begin{aligned}
p({\mu_u} \lvert \cdots)\propto & \prod_{i=1}^{N}\prod_{t=1}^T e^{-\frac{1}{2 \sigma_u^2} (y_{i,t}-\mu_u)^2 \mathds{1}_{U_{i,t}}(u)}  e^{-\frac{1}{2v} (\mu_u - m)^2} \\
 \propto & ~ e^{ -\frac{1}{2} \left[\mu_u^2(n_u/\sigma_u^2 + 1/v) - 2 \mu_u'(n_u {\bar{y}_u} / \sigma_u^2 + m/v)\right] },\label{aa1}
\end{aligned}
\end{equation*}
obtaining 
$$\mu_u \lvert \cdots \sim\mathcal{N}(\tilde{m} \tilde{v}, \tilde{v}),$$ 
where $\tilde{v} = (n_u/\sigma_k^2 + 1/v)^{-1}$ and $\tilde{m} = n_u {\bar{y}_u} / \sigma_k^2 + m/v$.
In addition, we have
\begin{equation*}
\begin{aligned}
p(\sigma_u^2 \lvert \cdots) \propto & (\sigma_u^2)^{-\frac{n_u}{2}} e^{-\frac{1}{2\sigma_u^2} \sum_i \sum_t (y_{i,t} - \mu_u)^2 \mathds{1}_{U_{i,t}}(u) } \\
& \times (\sigma_u^2)^{-a-1} e^{-\frac{b}{\sigma_u^2}} \\
& = (\sigma_u^2)^{-a-1-\frac{n_u}{2}} e^{-\frac{1}{\sigma_u^2} \left[ b + \frac{1}{2}\sum_i \sum_t (y_{i,t} - \mu_u)^2 \mathds{1}_{U_{i,t}}(u) \right]},
\end{aligned}
\end{equation*}
obtaining 
$$\sigma_u^2 \lvert \cdots \sim\mathcal{IG}\left(a+\frac{n_u}{2}, b + \frac{1}{2}\sum_i \sum_t (y_{i,t} - \mu_u)^2 \mathds{1}_{U_{i,t}}(u)\right).$$ 

\section*{Appendix D}

\newpage
\bibliographystyle{apalike}
\bibliography{reference}

\end{document}